\documentclass[reprint,amsmath,amssymb,aps,floatfix,showpacs,longbibliography,prl]{revtex4-1}

\usepackage[utf8]{inputenc} % Umlaute direkt verwenden, load before csquotes
\usepackage[babel]{csquotes} % context sensitive quotation, recommended with biblatex

\usepackage{graphicx}% Include figure files
\usepackage{dcolumn}% Align table columns on decimal point
\usepackage{bm}% bold math
\usepackage[]{units}	% example: \unit[val]{dim}
\usepackage{xcolor}
\usepackage{soul}
\usepackage{changes}
\usepackage{textcomp}
\usepackage{upgreek}
\usepackage{enumerate}
\usepackage{multirow} % columns in tables can span multiple rows
\usepackage{ulem}
\usepackage{amsmath}
\usepackage{bbold}
\usepackage{hyperref}% add hypertext capabilities
\usepackage[all]{hypcap}

\newcommand{\bra}[1]{\ensuremath{\left\langle#1\right|}}
\newcommand{\ket}[1]{\ensuremath{\left|#1\right\rangle}}

\begin{document}

\title{Electric-field-controlled cold dipolar collisions between trapped CH$_3$F molecules}%

\author{M. Koller}
\author{F. Jung}
\author{J. Phrompao}
\author{M. Zeppenfeld}
\author{I. M. Rabey}
\author{G. Rempe}

%\address{Max-Planck-Institut für Quantenoptik, Hans-Kopfermann-Strasse 1, 85748 Garching, Germany}
\affiliation{Max-Planck-Institut f\"ur Quantenoptik, Hans-Kopfermann-Strasse 1, 85748 Garching, Germany}
\date{\today}

\begin{abstract}
\noindent
Reaching high densities is a key step towards cold-collision experiments with polyatomic molecules. We use a cryofuge to load up to 2$\times10^7$ CH$_3$F molecules into a box-like electric trap, achieving densities up to 10$^7$/cm$^3$ at temperatures around 350\,mK where the elastic dipolar cross-section exceeds 7$\times$10$^{-12}$cm$^2$. We measure inelastic rate constants below 4$\times$10$^{-8}$cm$^3$/s and control these by tuning a homogeneous electric field that covers a large fraction of the trap volume. Comparison to ab-initio calculations gives excellent agreement with dipolar relaxation. Our techniques and findings are generic and immediately relevant for other cold-molecule collision experiments.
\end{abstract}

% Use showkeys class option if keyword display desired

\maketitle

\noindent
Polar molecules offer research opportunities that are not shared by other particles such as atoms \cite{Carr2009}. The strong and long-range electric dipole-dipole interaction in particular can affect quantum-chemical reaction pathways \cite{Herschbach2009,Ni2010,Quemener2012,Liu2021,Tobias2021}, can induce large-scale correlations and novel phases in molecular quantum gases \cite{Santos2000,Baranov2012,Wall2015,Yao2018a,Blackmore2019}, and can be the basis for a robust quantum-computing architecture \cite{DeMille2002,Yelin2006,Wei2011,Ni2018,Yu2019,Gregory2021}. Towards these applications, closed-shell symmetric-top molecules stand out as an ideal platform due to their simple rotational energy-level structure, favorable matrix elements for cycling transitions \cite{Zeppenfeld2009}, and linear response to an electric field \cite{Townes1975}. Together, these properties have allowed for direct cooling and trapping of the numerically largest samples of ultracold molecules to date \cite{Prehn2016}.

Further key requirements must be fulfilled to explore and leverage the dipole-dipole interaction between such molecules: First, observing dipolar collisions needs a high density combined with a long hold time. The latter can be accomplished by trapping the molecules \cite{Weinstein1998,Bethlem2000,Meerakker2005,Rieger2005,Sawyer2007, Englert2011,Akerman2017a,Mccarron2018,Williams2018}. Second, a high state purity is needed so that collision channels can be studied cleanly. This requires cooling the rotation of the molecules \cite{Maxwell2005,Glockner2015,Wu2016}. Also cooling the motion has the additional advantage that it increases the elastic cross-section and decreases the number of inelastic collision channels \cite{Bohn2009}. Third, manipulating the collision process calls for a suitable control technique that must be compatible with the aforementioned cooling and trapping \cite{Gorshkov2008,Quemener2011,Karman2018,Xie2020}. All demands have been met for ultracold dimers synthesized from laser-cooled \cite{Ospelkaus2010,Ni2010,Guo2018,Gregory2019,Yan2020,Li2021,Hu2021,Schindewolf2022} and directly cooled diatomic molecules \cite{Segev2019,Cheuk2020,Anderegg2021}. However, despite early attempts \cite{Sawyer2011} and recent advances \cite{Wu2017,Poel2018,Vilas2021}, collision studies with polyatomic molecules are still at a beginning.

Here we observe cold collisions between electrically trapped CH$_3$F molecules in a predominantly single rotational state. Moreover, we use a homogeneous electric field to tune the rate of inelastic two-body collisions. Excellent agreement between experimental data and a semi-classical model identifies the dominant loss mechanism to be dipolar relaxation \cite{Bohn2001} to untrapped rotational states. Understanding and controlling this mechanism is a sine-qua-non requirement for future thermalization and evaporative cooling experiments with molecules that can be decelerated, trapped and cooled, but are still far from the quantum regime.

\begin{figure}
    \centering
    \includegraphics[width=\columnwidth]{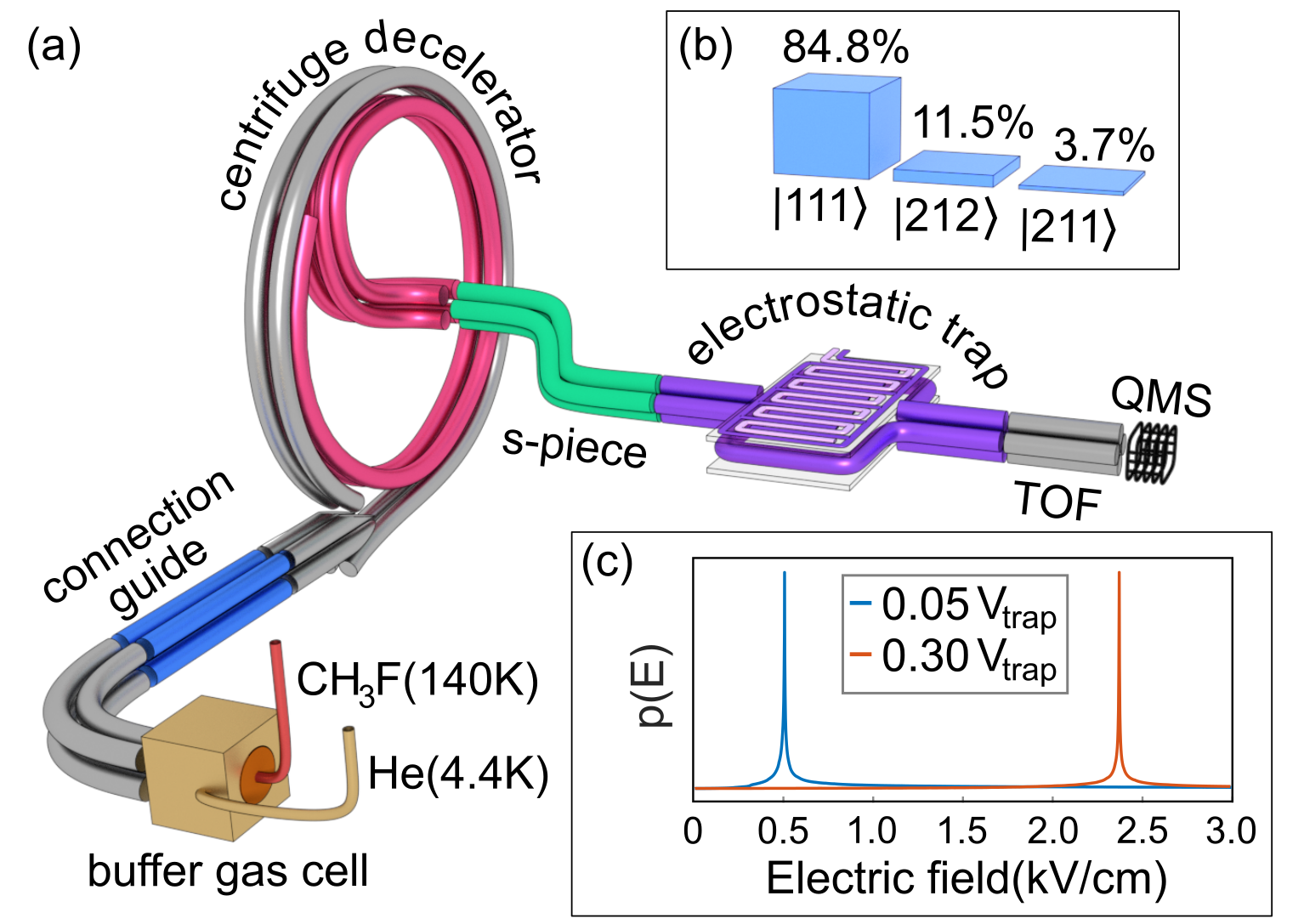}
    \caption{(Color online) Experimental setup. (a) CH$_3$F molecules are cooled in a cryogenic buffer-gas cell (Helium at 4.4\,K) and transferred to a centrifuge decelerator by an electric connection guide. A s-shaped guide acts as a velocity filter connecting the exit of the centrifuge to the inlet of the electric trap. A mass spectrometer at the end of a time-of-flight guide attached to the trap outlet detects the molecules. (b) Measured state distribution in the trap, given in the symmetric-top basis $\ket{\text{J,K,M}}$. (c) Simulated electric-field distribution \cite{Prehn2016} of the electrostatic trap for $E_{\text{c}}$$\,=\,$$0.50$\,kV/cm (blue) and $E_{\text{c}}$$\,=\,$$2.37$\,kV/cm (red). The field is homogeneous ($\leq$10\% relative deviation from $E_{\text{c}}$) over $\sim$50\% of the geometric trap volume.% , featuring a homogeneous electric field region $(E_{\text{c}}\pm0.1\cdot E_{\text{c}})$ that covers $\sim50\%$ of the trap volume.
    }
    \label{fig:setup}
\end{figure}

The starting point for our collision measurements is to create a high-density sample of CH$_3$F molecules confined in an electric trap. As a molecule source we employ our cryofuge \cite{Wu2017}, illustrated in Fig.\,\ref{fig:setup}(a), which combines cryogenic buffer-gas cooling \cite{Hutzler2012} with centrifuge deceleration \cite{Chervenkov2014} to produce a continuous, high-flux beam %(up to $\sim$10$^{10}$/s)
of trappable ($\leq$25\,m/s) molecules. The beam is velocity filtered by a sharply bent electric quadrupole \cite{Sommer2010} s-piece that connects the exit of the centrifuge with the input of the trap. This prevents the fast velocity tail of the guided molecules from reaching the trap. Loading is turned on and off by simultaneously switching the  guide connecting the cryogenic cell to the centrifuge decelerator and the s-piece between guiding and non-guiding configuration \cite{Sommer2010}. We tune the trap loading rate by varying the electric field of the connection guide between 2\,kV/cm for low flux and 20\,kV/cm for high flux ($\sim$10$^8$/s).

Our trap employs an electric multipole configuration that confines cold molecules in a box-like potential \cite{Englert2011}. It consists of a pair of microstructured capacitor plates, separated by 3\,mm, and a surrounding ring electrode. Alternating voltages $\pm V_{\text{trap}}$ are applied to the microstructure electrodes to provide strong confinement in the vertical direction, while the ring electrode with voltage $V_{\text{ring}}$ provides confinement in the other two horizontal dimensions. For all measurements presented here, the trapping and ring electrode voltages are fixed to $V_{\text{trap}}$$\,=\,$$1200$\,V and $V_{\text{ring}}$$\,=\,$$3V_{\text{trap}}$, resulting in a maximum trapping field of $E_{\text{trap}}$$\,=\,$$40$\,kV/cm. This confines molecules up to kinetic energies corresponding to $\sim$1\,K. A voltage difference applied to the capacitor plates creates a homogeneous electric control field $E_{\text{c}}$ that covers $\sim$50\% of the trapped molecule ensemble and is tuned between 0.50\,kV/cm and 2.37\,kV/cm, see Fig.\,\ref{fig:setup}(c). 

Molecules are unloaded from the trap via a time-of-flight (TOF) quadrupole guide, depicted in Fig.\,\ref{fig:setup}(a), that can be toggled on and off to measure the velocity distribution of the trapped sample (see Supplement). A quadrupole mass spectrometer (QMS) at the end of the guide detects the unloaded molecules, with the integrated signal being proportional to the density of trapped molecules, $n$. Its time evolution can be modeled by
\begin{equation}
    \frac{d}{dt}n(t)=\lambda(t)-\Gamma n(t)-k n^2(t),
    \label{eq:Densities}
\end{equation}
with $\lambda$ denoting the loading rate of molecules into the trap. Single-body loss from collisions with residual background gas, Majorana transitions for molecules passing through electric-field minima \cite{Kirste2009,Zeppenfeld2013}, or molecules leaving the trap through the input and output guides \cite{Englert2011,Zeppenfeld2013} are characterized by a single-body loss rate $\Gamma$. The density-dependent collision-induced trap losses are given by the two-body loss-rate coefficient $k$.

The standard approach to measure two-body loss is to observe a density-dependent non-exponential decay of the trapped sample, clearly distinct from a single-body exponential decay \cite{Segev2019}. We know, however, that in our trap $\Gamma$ depends strongly on the molecule velocity $v$ (proportional to $v^5$ for a linear Stark shift \cite{Zeppenfeld2013}). This causes deviations from a single-exponential decay even in the absence of collisions, i.e., in the limit of small density. However, for hold times less than 1\,s, single-body loss can be approximated by a single-exponential decay (see Supplement). A further problematic effect arises when tuning the loading rate. Despite all precautions, changing the connection guide voltage creates small changes in the velocity distribution of the trapped samples, affecting the single-body decay rate by an amount similar to that due to collisions. To counter these complications, we developed a new measurement scheme that is robust against small changes of $\Gamma$ (see Supplement), allowing us to extract a precise value for $k$ as described in the following.

\begin{figure}
    \centering
    \includegraphics{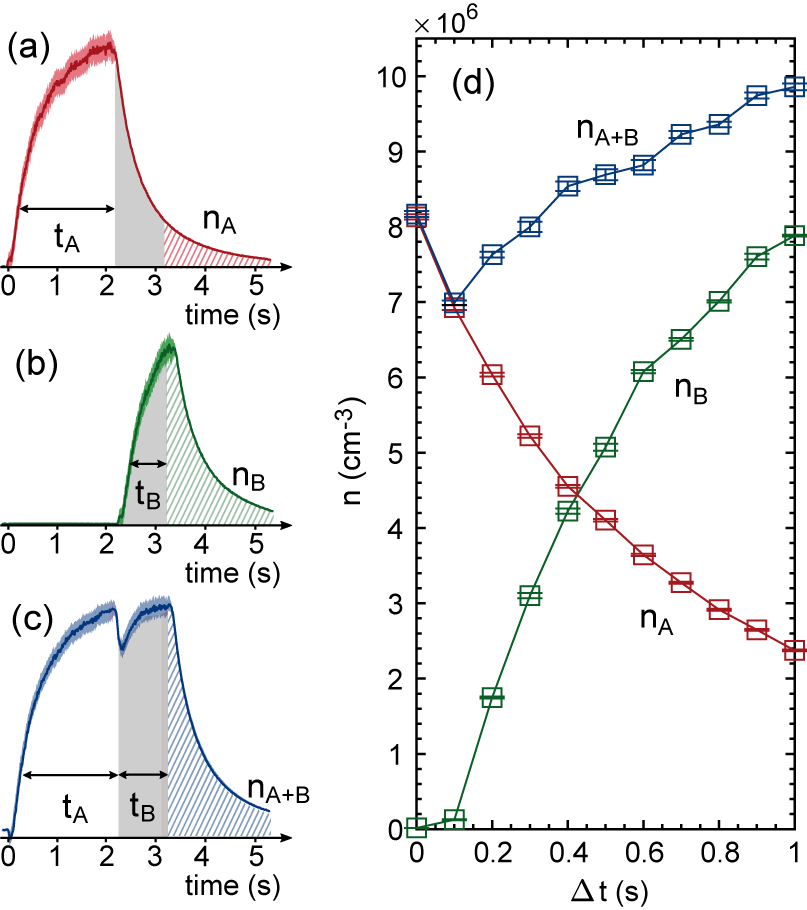}
    \caption{(Color online) Molecule signals. (a)-(c) Measurement sequences for samples A, B, and A+B. The grey areas depict the interaction time $\Delta t$, whereas the dashed areas illustrate the trap unloading signals which, when integrated, are proportional to the density of trapped molecules. (d) Density of trapped molecules as a function of the interaction time $\Delta t$ for the A, B, and A+B samples, recorded at $E_{\text{c}}$$\,=\,$$0.50$\,kV/cm.}
    \label{fig:scheme}
\end{figure}

Our measurement scheme combines the results of three distinct experiments with three molecular samples A, B, and A+B, with the trap effectively serving as a test tube. The two samples A and B are created independently and are separated by their loading times, as illustrated in Fig.\,\ref{fig:scheme}(a) and (b), respectively. Sample A is created by loading the trap for two seconds, reaching a steady state. At this time we stop the loading, wait 100\,ms for transient effects to disappear, and start the interaction period, $\Delta t$\,$\in$\,$[0,1]$\,s (grey-shaded area in Fig.\,\ref{fig:scheme}(a)-(c)), during which collisions occur. The density $n_A$ then evolves according to Eq.\,(\ref{eq:Densities}) with $\lambda_{\text{A}}$$\,=\,$$0$. Similarly, we create sample B by turning on the loading rate $\lambda_{\text{B}}$ for up to one second. We consider this to happen during the interaction period, but in the absence of sample A. For both samples, A and B, molecules are lost due to trap and collision losses between molecules, A-A collisions in sample A and B-B collisions in sample B. Finally, we create the third combined sample A+B by consecutively loading first sample A and then sample B, as illustrated in Fig.\,\ref{fig:scheme}(c). This allows for additional loss only by means of A-B collisions during the interaction period $\Delta t$, as both individual samples are independent in all other respects. Fig.\,\ref{fig:scheme}(d) shows the integrated trap unloading signals for the A, B, and A+B samples as a function of $\Delta t$. We then combine the densities measured in the three samples, $\delta n(t)$\,$=$\,$n_{\text{A}}(t)$\,$+$\,$n_{\text{B}}(t)$\,$-$\,$n_{\text{A+B}}(t)$, to extract the (positive) collision signal $\delta n$.

\begin{figure}
    \includegraphics{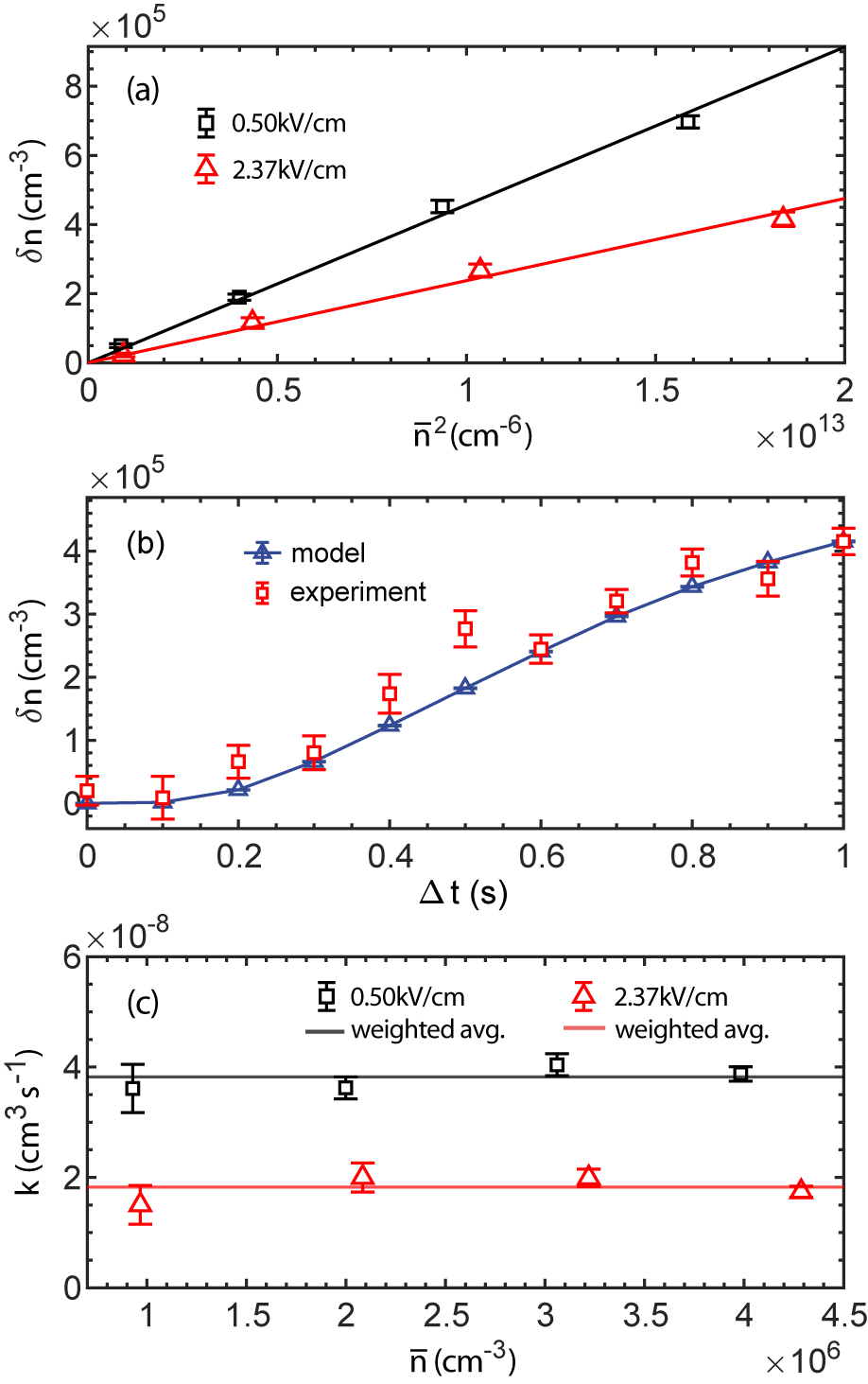}
    \caption{(Color online) Collision data. (a) Collision signal $\delta n$ due to A-B molecules interacting for $\Delta t=1$s as a function of $\bar{n}^2$, the product of $n_{\text{A}}(t)$ and $n_{\text{B}}(t)$ averaged over the interaction period, for two control fields. 
    (b) Solution of Eq.\,(\ref{eqn:nDifferential}) (blue triangles) fitted to the measured collision signal (red squares) at $\Delta t$$\,=\,$$1$\,s for $E_{\text{conn}}$$\,=\,$$20$\,kV/cm and $E_{\text{c}}$$\,=\,$$2.37$\,kV/cm. 
    (c) Two-body loss-rate coefficient $k$ plotted against $\bar{n}$, for two control fields. The solid lines depict the average of the measured data weighted by the respective error bars. Small changes in $k$ ($\leq 5\%$) might occur, as tuning the loading rate slightly alters the velocity distribution of the trapped ensembles.}
    \label{fig:N_AB}
\end{figure}

Experimentally, we tune the density of trapped CH$_3$F molecules by changing the electric field in the connection guide and record $\delta n$ as a function of $\bar{n}^2$, defined as the product of $n_{\text{A}}(t)$ and $n_{\text{B}}(t)$ averaged over $\Delta t$. Results for $\Delta t$$\,=\,$$1$\,s are displayed in Fig.\,\ref{fig:N_AB}(a) for $E_{\text{c}}$$\,=\,$$0.50$\,kV/cm and $E_{\text{c}}$$\,=\,$$2.37$\,kV/cm. The observed linear dependence of $\delta n$ on $\bar{n}^2$ proves, first, the existence of collisions (average collision energy $\sim$$k_\text{B}$$\times$0.4\,K) and, second, their nature as two-body loss process. The third observation refers to the clearly distinct slopes for the two control fields. This points to an electric-field dependence of $k$ that we investigate in the following.

To extract a precise value for $k$ we derive an expression for the time evolution of the collision signal $\delta n$ during the interaction period $\Delta t$, and fit this expression to the measured collision signal. By using Eq.\,(\ref{eq:Densities}) and the definition for $\delta n$ we obtain
\begin{equation} 
\begin{split}
\frac{d}{dt}\delta n(t) & = 2kn_{\text{A}}(t)n_{\text{B}}(t) - \delta n(t) \times\\
& \phantom{000} [\Gamma_{\delta n}+2k(n_{\text{A}}(t)+n_{\text{B}}(t))]+k(\delta n(t))^2,
\label{eqn:nDifferential} 
\end{split}
\end{equation}
where we use $\lambda_{\text{A+B}}$\,$=$\,$\lambda_{\text{A}}$\,+\,$\lambda_{\text{B}}$ and introduce the rate with which colliding molecules are lost in the A+B scenario via $\Gamma_{\delta n}\delta n=\Gamma_{\text{A}} n_{\text{A}}+\Gamma_{\text{B}} n_{\text{B}}-\Gamma_{\text{A+B}}n_{\text{A+B}}$.
The loss rates, $\Gamma_{\text{A}}$, $\Gamma_{\text{B}}$, $\Gamma_{\text{A+B}}$, and $\Gamma_{\delta n}$ are directly obtained from the measured data (see Supplement). With $k$ now being the only free parameter, we fit the solution of Eq.\,(\ref{eqn:nDifferential}) to the measured collision signal for $\Delta t$$\,=\,$$1$\,s. The result is displayed in Fig.\,\ref{fig:N_AB}(b) for $E_{\text{c}}$$\,=\,$$2.37$\,kV/cm and $E_{\text{conn}}$$\,=\,$$20$\,kV/cm (high input flux), yielding $k$\,$\simeq$\, 2$\times$10$^{-8}$cm$^3$/s. To test whether $k$ is a molecule-specific parameter that is independent of the density of trapped molecules, we perform a series of experiments with different control fields and molecule loading rates into the trap. Results are shown in Fig.\,\ref{fig:N_AB}(c) for $E_{\text{c}}$$\,=\,$$0.50$\,kV/cm and $E_{\text{c}}$$\,=\,$$2.37$\,kV/cm. For both electric fields, we observe a density independence for $k$, as expected for a molecule-specific parameter, but a clear electric-field dependence.

\begin{figure}
    \includegraphics{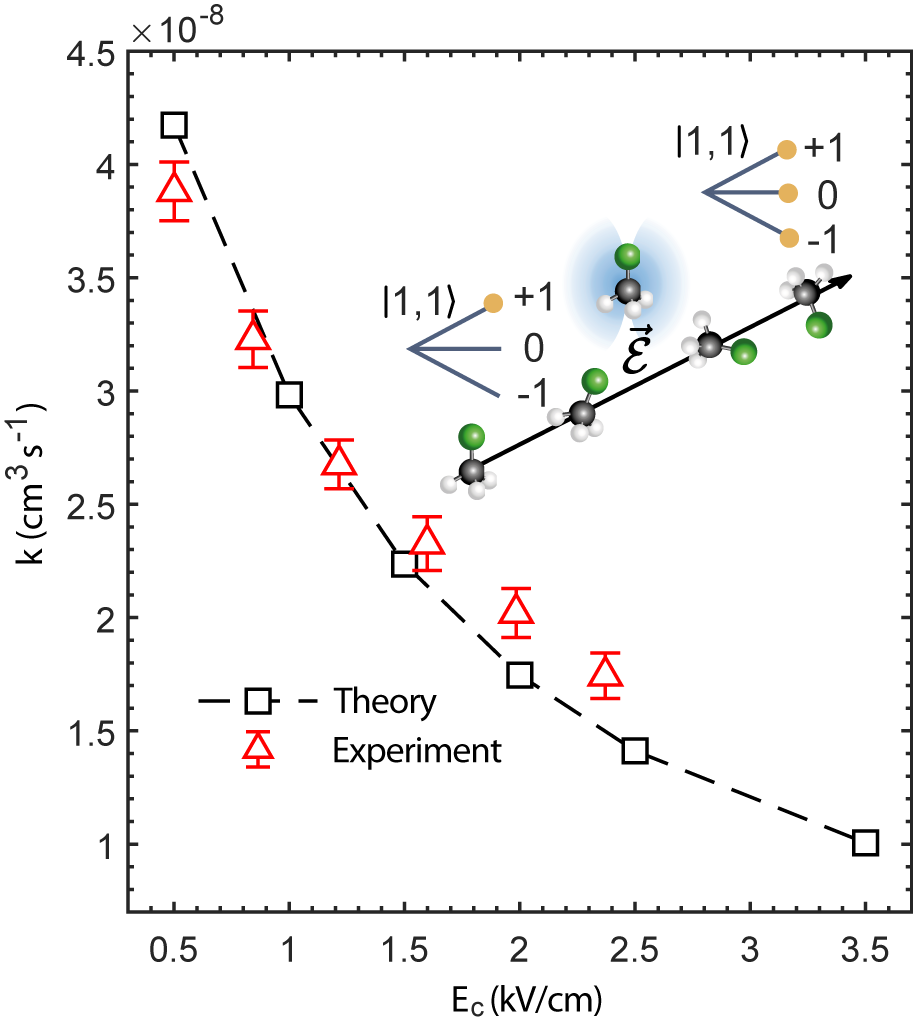}
    \caption{(Color online) Dipolar relaxation. Measured (red triangles) and calculated (black squares) two-body loss-rate coefficient $k$ of trapped molecules versus applied control field. The dashed line is a guide to the eye. The inset shows a schematic illustration of the collision process, with molecules being redistributed to lower M states. Information on the error budget of $k$ can be found in the Supplement.}
    \label{fig:kPlot}
\end{figure}

To investigate the latter in more detail, we tune the homogeneous control field $E_{\text{c}}$ to six different values between 0.50\,kV/cm and 2.37\,kV/cm and extract the corresponding values for $k$. The result is plotted in Fig.\,\ref{fig:kPlot}. Most striking is that $k$ reduces by more than a factor of two when $E_{\text{c}}$ increases from 0.50\,kV/cm to 2.37\,kV/cm. We interpret this observation as a clear signature of dipolar relaxation: The control field induces a Stark splitting and thereby an energy mismatch between the molecular internal-angular-momentum states that are coupled by the electric dipole-dipole interaction. During a non-adiabatic collision, the orientation of the dipole can change and population can be transferred from a trappable to a non-trappable state. The crucial point is that increasing the energy difference between the trappable and the non-trappable states reduces the probability for a non-adiabatic transfer. This simple picture predicts a decreasing loss-rate coefficient $k$ for increasing control field $E_{\text{c}}$, as observed in the experiment. An electric field has already been used to control chemical reactions \cite{Tobias2021} and evaporative cooling \cite{Li2021} of bialkali molecules, two applications distinct from our experiment. We note that for polyatomic molecules the electric field is a promising control parameter which should affect only the inelastic collisions, at least in our parameter regime. This should allow one to tune the ratio between elastic and inelastic collisions, a prerequisite for rethermalization and evaporative-cooling experiments \cite{Son2020,Schindewolf2022,Li2021}.

Beyond the qualitative picture outlined above, we now compare the data in Fig.\,\ref{fig:kPlot} with a quantitative ab-initio model. Towards this end, we consider elastic $k_{\text{el}}$ and inelastic $k_{\text{in}}$ contributions to $k$. Note that an elastic collision between two molecules leads to loss if the kinetic energy of one molecule after the collision is larger than the trap depth. The corresponding loss-rate coefficient is obtained from $k_{\text{el}}$\,$=$\,$\sigma^{\text{el}}_{\text{loss}}\,v_{\text{rel}}$ for a given relative velocity $v_{\text{rel}}$ with $\sigma^{\text{el}}_{\text{loss}}$ being the velocity-dependent elastic-loss cross-section. The latter is obtained from the differential elastic-collision cross-section $\frac{d\sigma}{d\Omega}(v_{\text{rel}},\theta)$ which is calculated using the semi-classical eikonal approximation \cite{Bohn2009}. The likelihood for a molecule to be lost from the trap after the collision, $P_{\text{loss}}(\theta,v_{\text{rel}})$ \cite{Wu2017}, is numerically determined from Monte-Carlo simulations that include the electric-field distribution and the velocity distribution of the molecules in the trap. When calculating $P_{\text{loss}}(\theta,v_{\text{rel}})$, we furthermore take into account that elastic collisions change the velocity distribution and thus the single-body loss rate $\Gamma$. By averaging $k_{\text{el}}$ over the relative-velocity distribution of the trapped molecules, we obtain $k_{\text{el}}$\,$=$\,$4.5\times 10^{-11}$cm$^{3}$/s for $E_{\text{c}}$\,$=$\,$2.37$\,kV/cm, which is about three orders of magnitude smaller than two-body loss-rate coefficients reported in Fig.\,\ref{fig:kPlot}. We emphasize that, although an elastic-collision process is unlikely to lead to loss, the elastic cross-section is estimated to be as large as $\sigma_{\text{el}}$\,$=$\,7.5$\times$10$^{-12}$cm$^2$. 

To calculate the total inelastic-loss-rate coefficient $k_{\text{in}}$, we first consider loss processes described by the short-range Langevin capture model \cite{Bell2009}. For a dipole moment of $d_{\text{avg}}$\,$=$\,$0.83$\,D, corresponding to the average for the measured state distribution in the trap, and a control field of $E_{\text{c}}$$\,=\,$$2.37$\,kV/cm, the Langevin loss-rate coefficient is obtained as $k_{\text{L}}$\,$=$\,6.8$\times$10$^{-10}$cm$^3$/s. This is larger than the above calculated $k_{\text{el}}$, but is again much smaller than the observed values displayed in Fig.\,\ref{fig:kPlot}, and is independent of $E_{\text{c}}$. Therefore, the Langevin model also fails to explain the observed losses.

To understand these, we now calculate the two-body dipolar-relaxation loss-rate coefficient $k_{\text{dd}}$. We do this by numerically solving the Schrödinger equation for a pair of molecules that move past each other along a fixed, straight trajectory in an electric field (see Supplement), as schematically illustrated in the inset of Fig.\,\ref{fig:kPlot}. The initial state vector, $\lvert\Psi(t$\,$=$\,$-\infty)\rangle$, takes into account the rotational-state distribution in the trap in the symmetric top basis $\ket{\text{J},\mp \text{K}, \pm \text{M}}$, with $\mp$\,$\text{K}$ chosen positive. Specifically, the molecules are statistically distributed over trappable states according to the buffer-gas cell temperature and the Stark shift. We measure the trapped state population via microwave depletion \cite{Zeppenfeld2012,Glockner2015a} to be $(84.8$\,$\pm$\,$0.7)\%$ in $\ket{1,1,1}$, $(7.3$\,$\pm$\,$0.7)\%$ in $\ket{2,1,2}$ and $(2.4$\,$\pm$\,$0.2)\%$ in $\ket{2,1,1}$, graphically illustrated in Fig.\,\ref{fig:setup}(b). The missing $(6.1$\,$\pm$\,$0.1)\%$ are distributed over higher-lying rotational levels with no single $\ket{\text{J,K,M}}$ state containing more than $1\%$ of the population. The dipole-dipole interaction redistributes the initial population over trappable and non-trappable states, and the state distribution after the collision process is obtained from $\lvert\Psi(t$\,$=$\,$+\infty)\rangle$. Summing the molecule population in non-trappable states over all possible trajectories and over the full solid angle then gives us the loss cross-section $\sigma_{\text{loss}}^{\text{dd}}(v_{\text{rel}},E_{\text{c}})$ for a given relative velocity of the colliding molecules and a given control field. We do not include the full electric-field distribution as this would only slightly alter $\sigma_{\text{loss}}^{\text{dd}}$ (by $\sim$\,10\%), but would increase the already long calculation time ($\sim$\,7\,months for the entire parameter space) more than tenfold. The loss-rate coefficient due to dipolar relaxation is now obtained from $k_{\text{dd}}(v_{\text{rel}},E_{\text{c}})$\,$=$\,$\sigma_{\text{loss}}^{\text{dd}}(v_{\text{rel}},E_{\text{c}})\,v_{\text{rel}}$, which we weight according to the measured relative-velocity distribution in the trap to get $k_{\text{dd}}(E_{\text{c}})$.

The sum of the elastic and inelastic contributions to $k$ are plotted and compared with experimental data in Fig.\,\ref{fig:kPlot} as a function of $E_{\text{c}}$. We use the calculated values for $k$ as an independent calibration of the molecule density, which we compare with the error-prone \cite{Wu2017} density value derived from the QMS signal. Thereby we find a scaling factor which we globally apply for all measurements presented here. Although this factor might affect the experimental value of $k$, the functional dependence $k(E_{\text{c}})$ as a molecule property is unaffected. We therefore attribute the observed losses to primarily (95\%) dipolar relaxation. This is confirmed by the fact that $k_{\text{dd}}$ is the only contribution with a pronounced electric-field dependence.

To conclude, we combined efficient cooling and deceleration with trapping of cold CH$_3$F molecules within an electric trap. We studied collisions in a clean and precisely controlled way, and changed the dipolar-relaxation loss rate by tuning the electric field. In the future we could add opto-electric Sisyphus cooling which has been applied in the same kind of trap to CH$_3$F \cite{Zeppenfeld2012} and H$_2$CO \cite{Prehn2016} for which temperatures as low as 420\,$\mu$K have been reached. Collision experiments with such cold molecules would benefit from a larger elastic cross-section and a smaller dipolar-relaxation loss rate, and thus could open up a route to quantum degeneracy.

This work was supported by Deutsche Forschungsgemeinschaft under Germany's excellence strategy via Munich Center for Quantum Science and Technology EXC-2111-390814868.

%\bibliographystyle{apsrev4-1}
%\bibliography{bibliography}


%\maketitle
\section{Supplement}

\section{Relative-velocity distribution of trapped molecules}\label{sec:RelVel}
\noindent
To calculate the two-body loss-rate coefficient for our system we consider contributions from the Langevin capture model \cite{Bell2009}, dipolar relaxation and elastic collisions, $k(v_{\text{rel}})=k_{\text{L}}(v_{\text{rel}})+k_{\text{dd}}(v_{\text{rel}})+k_{\text{el}}(v_{\text{rel}})$, with each of these components being a function of the relative velocity $v_{\text{rel}}$ of the colliding particles. In order to obtain a representative value for $k$ in our electric trap we average $k(v_{\text{rel}})$ over the molecules' relative-velocity distribution $D(v_{\text{rel}})$. The starting point to obtain $D(v_{\text{rel}})$ are time-of-flight measurements of the molecules unloaded from the trap. This enables us to extract the longitudinal-velocity distribution $D(v_z)$ from which we can deduce $D(v_{\text{tot}})$, the total-velocity distribution of the trapped molecules, to finally arrive at the relative-velocity distribution $D(v_{\text{rel}})$.

\begin{figure}
    \centering
    \includegraphics[width=1\hsize]{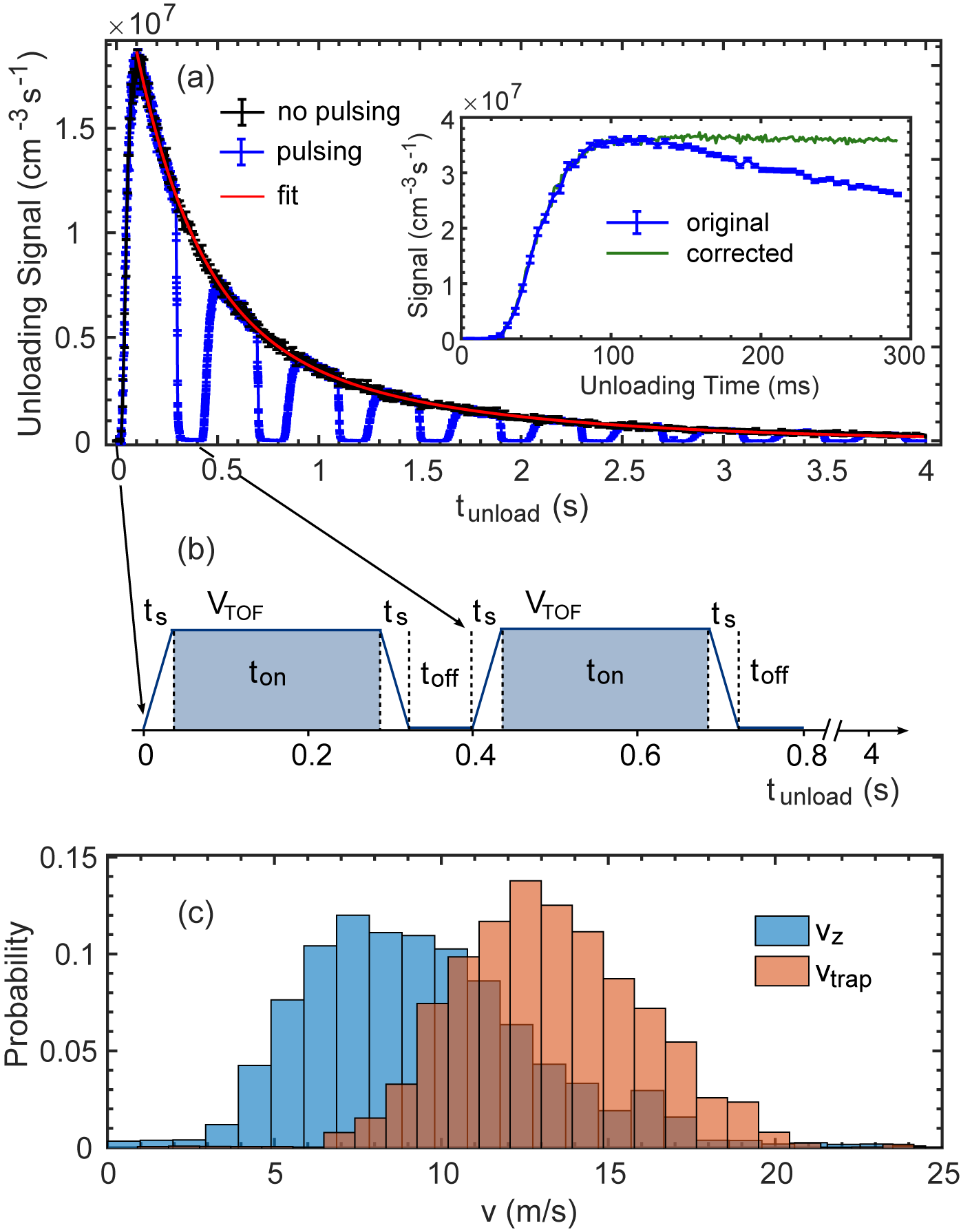}
    \caption{Longitudinal- and trapped-velocity distribution. (a) Trap unloading signal with the TOF-guide being in guiding configuration during the entire trap unloading time $t_{\text{unload}}$ is depicted in black and a double-exponential fit to the data in red. Toggling the TOF-guide repeatedly on and off during trap unloading is displayed in blue and the sum of the on-sequences is shown in the inset of (a). Here, blue shows the original data, while green is corrected for the double-exponential trap decay. (b) Illustration of the timing-sequence for the first two on-off blocks. (c) Longitudinal- (blue) and trapped- (orange) velocity distribution, for $E_{\text{c}}=0.50$\,kV/cm, obtained from the TOF-measurements.}
    \label{fig:vdist1}
\end{figure}

To record the time-of-flight measurements we toggle the TOF-guide between a guiding and a non-guiding configuration, while unloading the trap, as illustrated in blue in Fig.\,\ref{fig:vdist1} (a). Thereby we can extract the velocity of the molecules from the rising-edge signals for the entire trapped ensemble. In more detail, we apply a voltage of $V_{\text{TOF}}=3.6$\,kV in quadrupole configuration \cite{Sommer2010} to the TOF guide, connecting the trap exit to the detector, for $t_{\text{on}}=290$\,ms, as illustrated in Fig.\,\ref{fig:vdist1} (b). Consecutively, the TOF-guide is set to dipole configuration \cite{Sommer2010} for $t_{\text{off}}=90$\,ms, ensuring that no molecules can reach the detector during this time period. Including switching times of $t_{\text{s}}=10$\,ms this adds up to $400$\,ms for one on-off-sequence, which we repeat ten times during trap unloading to ensure that we obtain a representative velocity distribution of the trapped ensemble. The sum of the rising-edges is illustrated in blue in the inset of Fig.\,\ref{fig:vdist1} (a). We observe a rise in signal when switching the TOF-guide to guiding configuration. However, instead of reaching a steady state the signal decreases from $t_{\text{unload}}=100$\,ms onwards, as each block is overlaid by the trap decay, as can be seen in Fig.\,\ref{fig:vdist1} (a). To correct for this, we record the trap unloading signal with the TOF-guide being in guiding configuration throughout the entire trap unloading, depicted in black in Fig.\,\ref{fig:vdist1} (a), and fit it by a double-exponential function (red solid line). Using this fit we correct the sum of the rising-edges for the trap decay and obtain the data displayed in green in the inset of Fig.\,\ref{fig:vdist1} (a).
We can now use this data set, $S(t)$, to compute the longitudinal-velocity distribution \cite{Englert2013} according to
\begin{equation}
    D(v_z)=-\frac{dS(t)}{dv_z}=\frac{L}{v_z^2}\frac{dS(t)}{dt}
    \label{eqn:Dvz}
\end{equation}
with $L=51$\,cm being the length of the TOF guide. The resulting velocity distribution has a mean velocity of $\bar{v}_z=9.5$\,m/s and is depicted as a histogram in blue in Fig.\,\ref{fig:vdist1} (c). This specific measurement is performed for the A+B sample at a control field of $E_{\text{c}}=0.50$\,kV/cm, a trapping field in the connection guide of $E_{\text{conn}}=20$\,kV/cm and an interaction time of $\Delta t=0.5$\,s.
\begin{figure}
    \centering
    \includegraphics[width=1\hsize]{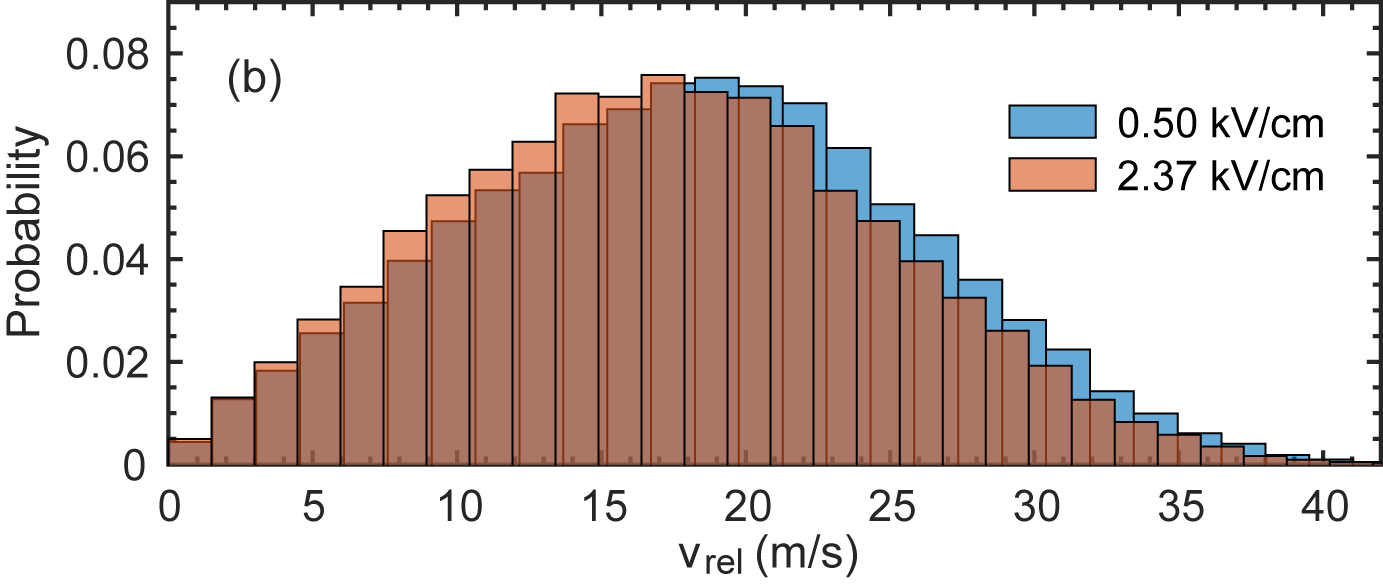}
    \caption{Relative-velocity distribution. Relative-velocity distributions of the A+B sample obtained from the TOF-measurements for a control field of $E_{\text{c}}=0.50$\,kV/cm (blue) and $E_{\text{c}}=2.37$\,kV/cm (orange) at an interaction time of $\Delta t=0.5$\,s.}
    \label{fig:vdist2}
\end{figure}

In a next step we utilize the longitudinal-velocity distribution $D(v_\text{z})$ to obtain the total-velocity distribution of the trapped molecules. Our electric trap provides uniform confinement in all three spatial dimensions, so that we can express the x-, y- and z-component of the total velocity $v_{\text{tot}}$ in spherical coordinates as
\begin{align}
\begin{split}
    v_x=v_{\text{tot}} \sin(\theta) \cos(\phi)\\
    v_y=v_{\text{tot}}\sin(\theta)\sin(\phi)\\
    v_z=v_{\text{tot}}\cos(\theta).\phantom{00.00}\\
\end{split}
\label{eqn:sphCoord}
\end{align}
With this, the total velocity can be obtained from the longitudinal velocity according to $v_{\text{tot}}=v_\text{z}/\cos(\theta)$, however, $\theta$ is an unknown parameter. To solve this issue, we use $D(v_\text{z})$ to determine $v_{\text{tot,max}}$, the maximum velocity a molecule can possess and still remain trapped. Therefore we extract the maximum longitudinal velocity $v_{\text{z,max}}$ from $D(v_\text{z})$, where the total velocity is solely given by its z-component, $v_{\text{tot,max}}=v_{\text{z,max}}$. For any measured $v_{\text{z}}$, the total velocity can now take any value between $v_{\text{z}}$ and $v_{\text{z,max}}=v_{\text{tot,max}}$. The latter corresponds to an upper bound on the inclination $\theta$, which can be computed using Eq.\,\ref{eqn:sphCoord} as
\begin{equation}
    \theta_{\text{max}}(v_{\text{z}})=\arctan\left(\sqrt{\frac{v_{\text{z,max}}^2-v_z^2}{v_z^2}}\right).
    \label{eqn:thetaMax}
\end{equation}
To calculate a distribution of possible values for $v_{\text{tot}}$ for one given $v_{\text{z}}$, we thus sample $\theta$ uniformly from the interval $[0,\theta_{\text{max}}(v_{\text{z}})]$ according to $v_{\text{tot}}=v_\text{z}/\cos(\theta)$. By taking the mean value of each distribution of total velocities we obtain one $v_{\text{tot}}$-value for a given $v_\text{z}$. By performing this procedure for the whole distribution $D(v_\text{z})$ we finally obtain $D(v_{\text{tot}})$, the total-velocity distribution of the molecules in our electric trap in the absence of an applied control field. To include the control field we utilize the Stark effect $E_{\text{pot}}=-d_{\text{avg}} E_{\text{c}}$, with $d_{\text{avg}}$ being the dipole moment averaged over the measured state distribution in the trap. Thereby we obtain the kinetic energy of a given molecule of mass $m$ according to $E_{\text{kin}}=\frac{1}{2}mv_{\text{tot}}^2-E_{\text{pot}}$, which we utilize to calculate the velocity distribution $D(v_{\text{trap}})$. Fig.\,\ref{fig:vdist1} (c) shows an example velocity distribution of trapped molecules in the presence of an applied control field $E_{\text{c}}=0.50$\,kV/cm for the A sample and an interaction time of $\Delta t=1$\,s (shown in orange).

In the last step we calculate the relative-velocity distribution $D(v_{\text{rel}})$ using $D(v_{\text{trap}})$, where the result is depicted in Fig.\,\ref{fig:vdist2} for the two control fields $E_{\text{c}}=2.37$\,kV/cm (orange) and $E_{\text{c}}=0.50$\,kV/cm (blue). As expected, increasing the control field leads to a decrease of the kinetic energy of the molecules in the trap, such that the two samples illustrated in Fig.\,\ref{fig:vdist2} differ by $1$\,m/s in their average velocity.

\section{Semi-classical model to calculate dipolar relaxation}

In this section we provide additional information on the numerical calculation of the electric-field-dependent two-body loss-rate coefficient $k_{\text{dd}}(E_{\text{c}})$ of dipolar relaxation. The Hamiltonian describing the system is given by \cite{Wall2013a}, 
\begin{equation}
    \hat{H}=\hat{H}_{s}+\hat{H}_{dd}
    \label{eqn:HS_Suppl}
\end{equation}
where the interaction of the molecules with an external electric field is expressed by the Stark Hamiltonian
\begin{equation}
    \hat{H}_{s}=-(\boldsymbol{\hat{d}}_1+\boldsymbol{\hat{d}}_2)\cdot\boldsymbol{E}
\end{equation}
with dipole moment operator $\boldsymbol{\hat{d}}_i$ of particle $i=1,2$ and electric field $\boldsymbol{E}=E_{\text{c}}\boldsymbol{e_\text{z}}$, defining the z-axis of the system. By applying the Wigner-Eckart theorem \cite{Edmonds1996} we evaluate the matrix elements of $\hat{H}_{s,i}=-\boldsymbol{\hat{d}_{i}}\cdot\boldsymbol{E}$ in the single-particle symmetric top basis $\ket{J,K,M}$, as 

\bigskip
$\bra{J_i',K_i',M_i'}\hat{H}_{s,i}\ket{J_i,K_i,M_i}=$
\begin{align}
\begin{split}
-dE_{\text{c}}(-1)^{M_i'-K_i'}\sqrt{(2J_i'+1)(2J_i+1)}\times\\
    \left( \begin{array}{rrr}
    J_i' & 1 & J_i \\ 
    -K_i' & 0 & K_i \\ 
    \end{array}\right)
    \left( \begin{array}{rrr}
    J_i' & 1 & J_i \\ 
    -M_i' & 0 & M_i \\ 
    \end{array}\right).
\end{split}
\end{align}
Following the standard definition \cite{Townes1975}, $J$ is the total angular momentum, $K$ its projection onto the molecule's symmetry axis and $M$ the projection of $J$ on the electric-field axis. Using the results for the single-particle basis, we obtain the electric field response of the two-particle state
$\ket{J_1,K_1,M_1} \otimes \ket{J_2,K_2,M_2}=\ket{J_1,K_1,M_1,J_2,K_2,M_2}$ according to 
\begin{equation}
    \hat{H}_{s}=\hat{H}_{\text{s,1}}\otimes\mathbb{1}+\mathbb{1}\otimes\hat{H}_{s,2}.
\end{equation}
Besides the interaction with an external electric field, we also have to consider the dipole-dipole interaction, described by \cite{Wall2013a} 
\begin{equation}
    \hat{H}_{\text{dd}}=\frac{\boldsymbol{\hat{d}}_1\cdot\boldsymbol{\hat{d}}_2-3(\boldsymbol{\hat{d}}_2\cdot\boldsymbol{e}_r)(\boldsymbol{e}_r\cdot\boldsymbol{\hat{d}}_1)}{4\pi\epsilon_0|\boldsymbol{r}(t)|^3}
    \label{eqn:Hdd}
\end{equation}
with $\epsilon_0$ being the vacuum permittivity, $\boldsymbol{e}_r$ the unit vector pointing from molecule 1 to molecule 2 and $\boldsymbol{r}(t)$ the time-dependent distance between the two molecules. We evaluate the matrix elements of $\hat{H}_{dd}$ as \cite{Wall2013a} 

\bigskip
$\bra{J_1',K_1',M_1',J_2',K_2',M_2'}\hat{H}_{\text{dd}}\ket{J_1,K_1,M_1,J_2,K_2,M_2}=$
\begin{align}
\begin{split}
    -\sqrt{30}\frac{d^2}{4\pi\epsilon_0|\boldsymbol{r}(t)|^3}(-1)^{M_1'-K_1'+M_2'-K_2'}\times
    \phantom{00000}\\
    \sqrt{(2J_1'+1)(2J_1+1)(2J_2'+1)(2J_2+1)}\times\phantom{00000}\\
    \sum_{p=-2}^2(-1)^pC_{-p}^{(2)}(\theta,\phi)\sum_{m=-1}^1(-1)^p
    \left( \begin{array}{rrr}
    1 & 1\phantom{00} & 2 \\ 
    m & p-m & -p \\ 
    \end{array}\right)\times\\
    \left( \begin{array}{rrr}
    J_1' & 1 & J_1 \\ 
    -K_1' & 0 & K_1 \\ 
    \end{array}\right)       
    \left( \begin{array}{rrr}
     J_1' & 1 &  J_1 \\ 
    -M_1' & m & M_1 \\ 
    \end{array}\right)   
     \left( \begin{array}{rrr}
    J_2' & 1 & J_2 \\ 
    -K_2' & 0 & K_2 \\ 
    \end{array}\right)\times\\     
    \left( \begin{array}{rrr}
     J_2' & 1\phantom{00} &  J_2 \\ 
    -M_2' & p-m & M_2 \\
    \end{array}\right)\phantom{000000000000000000000000000}     
\end{split}
\label{eqn:ME_Hdd}
\end{align}

where $C_{-p}^{(2)}(\theta,\phi)$ are the unnormalised spherical harmonics. The dipole-dipole coupling redistributes the initially trapped population of the colliding molecules over trappable and untrappable rotational states, which is known as dipolar relaxation \cite{Bohn2001}. 

\begin{figure}
    \centering
    \includegraphics{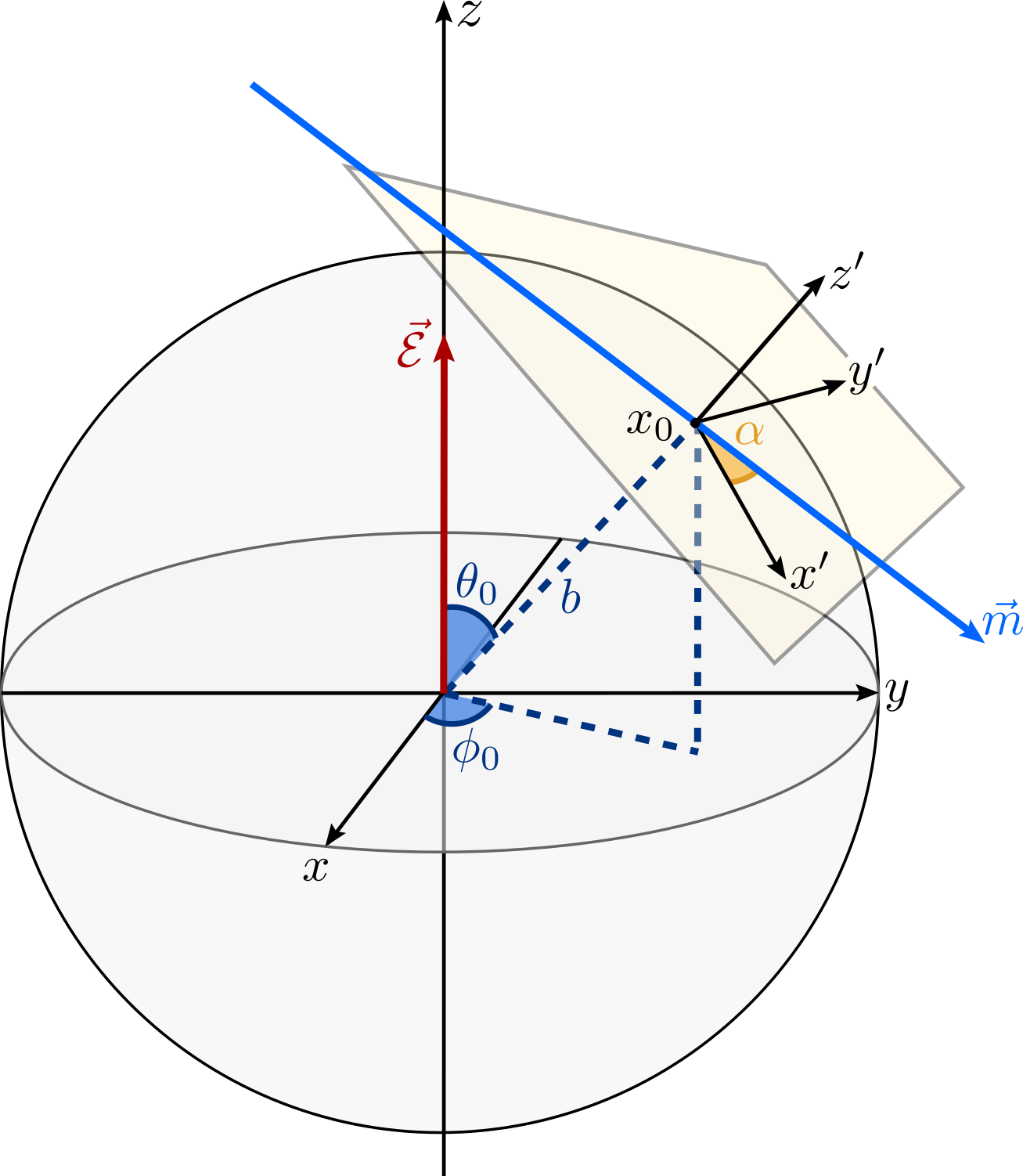}
    \caption{Molecule trajectory. Schematic illustration of the molecule trajectory in the collision process. Molecule 1 is fixed at the origin of the sphere, while molecule 2 is moving on the trajectory $\boldsymbol{m}$. The externally applied electric field defines the z-axis of the system.}
    \label{fig:sphere}
\end{figure}

We account for the molecules' movement in the trap with the time-dependent inter-particle distance $\boldsymbol{r}(t)$. To derive an expression for $\boldsymbol{r}(t)$ we start with fixing the position of molecule $1$ at the origin of the coordinate system, which is defined by the electric field pointing in z-direction $\boldsymbol{E}=E_{\text{c}}\boldsymbol{e_z}$, as depicted in Fig.\,\ref{fig:sphere}. Molecule 2 is traveling past Molecule $1$ on an arbitrary straight-line trajectory, described by the unit vector $\boldsymbol{\hat{m}}$, which can be related to the inter-particle distance via $\boldsymbol{r}(t)=b\boldsymbol{m_0}+v_{\text{rel}}t\boldsymbol{\hat{m}}$. Here, $v_\text{rel}$ is the relative velocity of the two molecules, $b$ the impact parameter and $t$ the time with $t\in[-\infty,\infty]$. The minimal distance between the two molecules is given by the impact parameter $b$ at time $t=0$, where the position of molecule 2 is $\boldsymbol{m_0}$. Within the plane perpendicular to $b\boldsymbol{m_0}$, the trajectory of molecule 2 can be in any direction through the point $\boldsymbol{m_0}$. Therefore $\alpha$, the angle between $x'-$axis and the unit vector $\boldsymbol{\hat{m}}$ along the trajectory, shown in Fig.\,\ref{fig:sphere}, can take any value between $0$ and $2\pi$. The orientation of this plane with respect to the z-axis is given by $\theta_0$, the orientation with respect to the x-axis by $\phi_0$. With this we can describe the movement of molecule 2 with time-dependent spherical coordinates, given by 

\begin{equation}
    r=\sqrt{b^2+v_{\text{rel}}^2t^2}
\end{equation}

\begin{equation}
    \theta=\cos^{-1}\left(\frac{b\cos{\theta_0}-v_{\text{rel}}t\cos{\alpha}\sin{\theta_0}}{\sqrt{b^2+v_{\text{rel}}^2t^2}}\right)
    \label{eqn:cos}
\end{equation}

\begin{equation}
    \phantom{0000}\phi=\tan^{-1}\left(\frac{v_{\text{rel}}t\sin{\alpha}}{b\sin{\theta_0}+v_{\text{rel}}t\cos{\alpha}\cos{\theta_0}}\right),
    \label{eqn:sin}
\end{equation}

\bigskip
which we insert into Eq.\,\ref{eqn:ME_Hdd}. As the electric field $\boldsymbol{E}=E_{\text{c}}\boldsymbol{e_z}$ defines the z-axis in our system we made use of the resulting cylindrical symmetry of the problem and set $\phi_0=0$ in Eq.\,\ref{eqn:cos} and \ref{eqn:sin} as it only adds a phase factor of $2\pi$.

An important assumption in our model is that we consider the trajectory of molecule $2$ to be a straight-line trajectory that is not altered by the dipole-dipole potential 
\begin{equation}
    V_{dd}(r,t)=-\frac{d^2}{4\pi\epsilon_0|\boldsymbol{r}(t)|^3}.
\label{eqn:Vdd}
\end{equation}
We will justify this in the following by exemplarily considering a pair of molecules colliding with $v_{\text{rel}}=19.8$\,m/s, the mean relative velocity of the A+B sample for $E_{\text{c}}=0.50$\,kV/cm and $\Delta t=0.5$\,s, at an impact parameter of $b=\sqrt{\sigma_{\text{loss}}^{\text{dd}}/\pi}=1.83\times10^{-8}$\,m. As introduced in the previous paragraph, molecule 1 is fixed at the origin of the coordinate system so that we only have to compare the momentum of molecule 2, $p_{\text{mol}}=m v_{\text{rel}}$, with the momentum transfer induced by the dipole-dipole potential
\begin{equation}
    p_{\text{dd}}=\int_{-\infty}^{\infty} \frac{dV_{\text{dd}}(r,t)}{dr}dt,
\label{eqn:pdd}
\end{equation}
to calculate the deflection of the trajectory of molecule 2. To give an upper bound on the deflection we assume $p_{\text{dd}}$ to be perpendicular to $p_{\text{mol}}$, resulting in a deflection of the trajectory of molecule 2 by only $\sim 0.7^{\circ}$. This shows that due to the large mean collision energy of $\bar{E}_{\text{coll}}=k_\text{B}\cdot0.4K$ in our trap it is justified to consider the trajectory of molecule 2 as a straight-line trajectory.

At this point we want to note that in principle there is a third contribution to the Hamiltonian $\hat{H}$, introduced in Eq.\,\ref{eqn:HS_Suppl}; the Hamiltonian $\hat{H}_{\text{rot}}$, describing the rotational-energy structure of the molecule \cite{Bohn2001}. However, as we will show in the following, we can neglect this term and thereby significantly reduce the dimensions of $\hat{H}$ to save computation time. Dipolar-relaxation-induced population transfer between a pair of states with energy separation $\Delta$ is only possible if the transition is non-adiabatic, where the likelihood for such a transfer decreases with increasing energy separation. Following Zeppenfeld \cite{Zeppenfeld2017}, we can estimate an upper limit $\Delta_{\text{max}}/h\approx15$\,GHz using Eq.\,\ref{eqn:ME_Hdd} with $\phi=0$ and $\theta=\pi/2$ for a collision-induced population transfer in our system. This is significantly smaller than $\sim 102$\,GHz rotational splitting \cite{Wu2016} between the only significantly populated (J,K)-manifolds, (1,1) and (2,1), in our experiment. Therefore we can perform separate calculations for these two manifolds and weight them according to the state distribution in the trap. In addition, we also need to take into account exchange collisions between the (1,1)- and the (2,1)-manifold, as the possible energy mismatch between the initial and the final state of the two-particle system is only given by the Stark splitting and therefore on the order of a few GHz. To get a more intuitive picture for this specific collision process we consider an example, where molecule 1 with initial state $\ket{J_1'=1,K_1'=1,M_1'=1}$ collides with molecule $2$ with initial state $\ket{J_2'=2,K_2'=1,M_2'=1}$. We assume the dipole-dipole interaction to redistribute the population of molecule 1 and 2 to the non-trappable states $\ket{J_1=2,K_1=1,M_1=0}$ and $\ket{J_2=1,K_2=1,M_2=0}$, respectively. In this situation the energy splitting between the  states of the molecules before and after the collision is just a few GHz as it is given by the Stark shift of the two-particle system. In contrast, if $\hat{H}_{\text{dd}}$ transfers the population of molecule 1 and 2 to $\ket{J_1=2,K_1=1,M_1=0}$ and $\ket{J_2=2,K_2=1,M_2=0}$, respectively, the energy separation between the total energy splitting between the states of the molecules before and after the collision is on the order $\sim 102$\,GHz and therefore highly unlikely.

With this in place we can now determine the two-body loss-rate coefficient of dipolar relaxation $k_{\text{dd}}$. Utilizing the measured state population (see main text) we obtain the initial state vector $\ket{\Psi(t=-\infty)}$ which we use to solve the Schrödinger equation for the Hamiltionian $\hat{H}$ (see Eq.\,\ref{eqn:HS_Suppl}),
\begin{equation}
    i\hbar\frac{d}{dt}\ket{\Psi(t)}={\hat{H}\ket{\Psi(t)}}.
\end{equation}
Thereby we can obtain the state population after the collision process \cite{Zeppenfeld2017} in untrapped states $\ket{\Psi_{\text{hfs}}(t=\infty,\theta_0,b,\alpha)}$, which we use to compute the loss cross-section due to dipolar-relaxation $\sigma_{\text{loss}}^{\text{dd}}$ for a given control field $E_{\text{c}}$ and relative velocity $v_{\text{rel}}$ according to
\begin{align}
\begin{split}
    \sigma_{\text{loss}}^{\text{dd}}(E_{\text{c}},v_{\text{rel}})=
        \phantom{00000000000000000000000000}\\
        2\pi\int_{0}^{2\pi} \int_{0}^{\infty}  \int_{0}^{\pi}|\Psi_{\text{hfs}}(t=\infty)|^2 \sin(\theta_0)\ d\theta_0 \ b  db \ d\alpha.
\end{split}
\label{eqn:intloss}
\end{align}

\begin{figure}[h!]
\centering
\includegraphics[width=1\hsize]{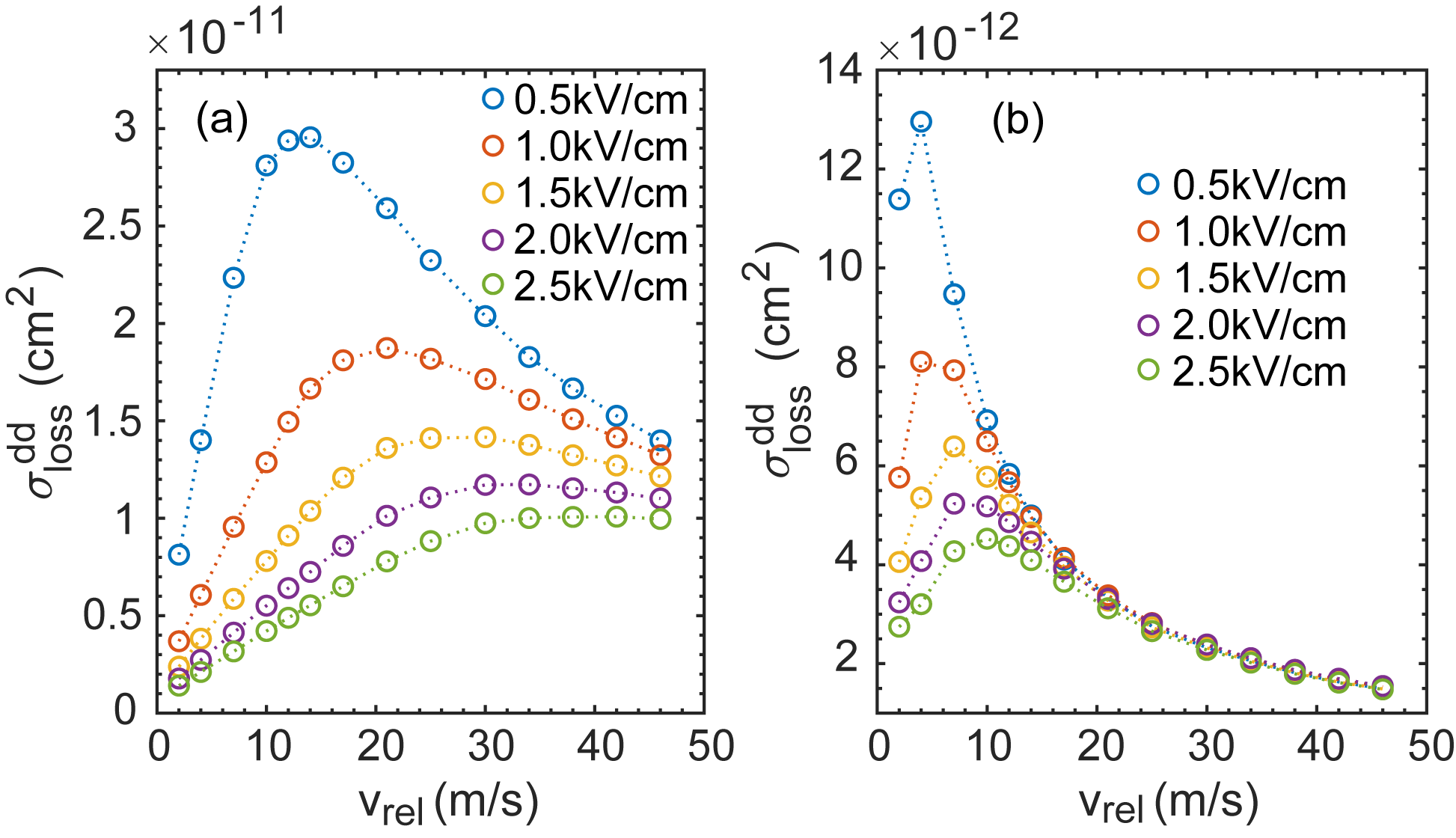}
\caption{Loss cross-section of dipolar relaxation. (a) and (b) show the loss cross-sections for the (1,1)- and (2,1)-manifold as a function of the relative velocity $v_{\text{rel}}$ of the colliding molecules, respectively. Both plots include data for control fields from $E_{\text{c}}=0.5$\,kV/cm to $E_{\text{c}}=2.5$\,kV/cm.}
\label{fig:sigmavn}
\end{figure}
\noindent
As the applied electric field does not only define the z-axis of our system but also induces a cylindrical symmetry to the problem we directly evaluated the integral over $\phi_0$ in Eq.\,\ref{eqn:intloss} since it just adds a phase factor of $2\pi$. In contrast, for each value of  $\theta_0$, $\alpha$ and $b$ we have to solve the Schrödinger equation to compute $\sigma_{\text{loss}}^{\text{dd}}(E_{\text{c}},v_{\text{rel}})$. We found the best compromise between computation time and accuracy of the calculation results using step sizes of $20^{\circ}$ for both $\theta_0$ and $\alpha$ and a step size of $db=5\times10^{-10}$m for the impact parameter. We want to note that finer step sizes only provide improvements to the accuracy in the low single-digit percent range. We calculate the loss cross-section due to dipolar-relaxation for five control fields $E_{\text{c}}$\,$=$\,$[0.5,1.0,1.5,2.0,2.5]$\,kV/cm and 14 different relative velocities, covering the entire relative-velocity distribution in our trap ($v_{\text{rel}}$\,$=$\,$[2,4,7,10,12,14,17,21,25,30,34,38,42,46]$\,m/s).

Fig.\,\ref{fig:sigmavn} (a) and (b) illustrate $\sigma_{\text{loss}}^{\text{dd}}$ for the (1,1)- and the (2,1)-manifold as a function of the relative velocity for the five control fields, respectively. For a molecule to be lost from the trap due to dipolar relaxation, a non-adiabatic transition from a trappable to a non-trappable state has to occur. According to Landau-Zener theory, the transition rate $\gamma$, which is proportional to the relative velocity, has to be larger than the energy separation between the states coupled by the dipole-dipole interaction. Therefore we see an increase in the loss cross-section in Fig.\,\ref{fig:sigmavn} (a) and (b) for increasing $v_{\text{rel}}$ until $\gamma$ is clearly larger than the Stark splitting of the states coupled by $\hat{H}_{\text{dd}}$. From this point onwards, the loss cross-section reduces for increasing relative velocities.

As the energy separation between the states is given by the Stark shift in our system, it can be controlled by tuning an external electric field. The result is visible in Fig.\,\ref{fig:sigmavn} (a) and (b), showing that a larger electric field causes a larger energy separation between the states coupled by $\hat{H}_{\text{dd}}$, so that the losses due to dipolar relaxation are suppressed and larger relative velocities are required for a non-adiabatic transition.

\section{Langevin capture model}

Besides collisional loss caused by the long-range dipole-dipole interaction we assume any further inelastic two-body loss to be accounted for by the Langevin capture model \cite{Bell2009}. The associated loss-rate coefficient can be obtained as \cite{Zhang2017}
\begin{equation} 
k_{\text{L}}(v_{\text{rel}})=3\pi v_{\text{rel}}\left(\frac{d_{\text{avg}}^2}{4\pi\mu\epsilon_0 v_{\text{rel}}^2}\right) ^{2/3},
\label{eqn:langevin} 
\end{equation}
with $d_{\text{avg}}$ being the dipole moment averaged over the rotational-state distribution in the trap, $v_{\text{rel}}$ the relative velocity of the colliding molecules, $\mu$ the reduced mass and $\epsilon_0$ the vacuum permittivity. We average $k_{\text{L}}(v_{\text{rel}})$ over the relative-velocity distribution $D(v_{\text{rel}})$ to obtain the Langevin loss-rate coefficient $k_{\text{L}}$ for the trapped ensemble at a given control field $E_{\text{c}}$. As can be seen from Eq.\,\ref{eqn:langevin} the Langevin capture model does not show an explicit electric-field dependence. However, as the control field alters the trapped-velocity distribution, $k_{\text{L}}$ is indirectly affected by $E_{\text{c}}$. By computing the Langevin loss-rate coefficient for the two control fields $E_{\text{c}}=0.50$\,kV/cm and $E_{\text{c}}=2.37$\,kV/cm as $k_{\text{L}}=6.67\times10^{-10}$cm$^3$/s and $k_{\text{L}}=6.82\times10^{-10}$cm$^3$/s, we confirm that the impact of $E_{\text{c}}$ on $k_{\text{L}}$ is small. More importantly, we want to emphasize that the Langevin loss-rate coefficient is roughly two orders of magnitude smaller than the loss-rate coefficient due to dipolar relaxation $k_{\text{dd}}$.

\section{Extraction of the two-body loss-rate coefficient $k$}

In this section we derive an expression for the additional loss $\delta n$, caused by the interaction of the A and the B sample, which we can fit to the collision measurements presented in the main text to obtain a value for the two-body loss-rate coefficient $k$. Our starting point is the time evolution of the densities of the individual trapped samples, given by 
\begin{equation}
    \dot{n}_{\text{x}}=\lambda_{\text{x}}-\Gamma_{\text{x}} n_{\text{x}}-k n_{\text{x}}^2,
    \label{eqn:Densities}
\end{equation}
where x can be the A,B or A+B sample, $\lambda_{\text{x}}$ is the loading rate of molecules into the trap and $\Gamma_{\text{x}}$ the single-body loss rate. For non-interacting samples, or equivalently $k=0$, $\Gamma_{\text{A+B}}$ is given by the weighted sum of the single-body loss rates of the A and the B sample as
\begin{equation}
    \Gamma_{\text{A+B}}=\frac{\Gamma_{\text{A}} n_{\text{A}}+\Gamma_{\text{B}} n_{\text{B}}}{n_{\text{A+B}}}.
\end{equation}
However, for $k\neq0$ molecules are lost in the A+B scenario due to the interaction of the A and the B sample, according to $\delta n=n_{\text{A}}+n_{\text{B}}-n_{\text{A+B}}$. As a consequence the single-body loss rate of the A+B sample can change for $k\neq0$ due to the energy dependence of $\Gamma$ in our system. However, by introducing $\Gamma_{\delta n}$, the single-body loss rate of $\delta n$, we can define a corrected expression for $\Gamma_{\text{A+B}}$, taking into account the losses due to A-B collisions
\begin{equation}
    \Gamma_{\text{A+B}}\approx\frac{\Gamma_{\text{A}} n_{\text{A}}+\Gamma_{\text{B}} n_{\text{B}}-\Gamma_{\delta n}\delta n}{n_{\text{A+B}}}.
    \label{eqn:Gamma_n}
\end{equation}
With this we have everything in place to derive the time evolution of the additional loss $\delta n$. Therefore we take the time derivative of $\delta n=n_{\text{A}}+n_{\text{B}}-n_{\text{A+B}}$ and insert the time evolution of the individual samples, given by Eq.\,\ref{eqn:Densities}. In addition we exploit the fact that $\lambda_{\text{A+B}}=\lambda_{\text{A}}+\lambda_{\text{B}}$ and use Eq.\,\ref{eqn:Gamma_n} to finally arrive at 
\begin{equation} 
\dot\delta n = 2kn_{\text{A}}n_{\text{B}} - \delta n [\Gamma_{\delta n}+2k(n_{\text{A}}+n_{\text{B}})]+k(\delta n)^2.
\label{eqn:nDifferentialSuppl} 
\end{equation}
As we will see later, the contribution of the quadratic term $k (\delta n)^2$ is small, so that we neglect it at first, solve the linear part of the differential equation analytically and include the quadratic contribution in a pertubative approach. Therefore we insert the ansatz $\delta n=\delta n_0 + \epsilon\delta n_1$ into Eq.\,\ref{eqn:nDifferential} and order the terms according to the power of $\epsilon$, resulting in the differential equations
\begin{equation} 
\dot\delta n_0 = 2kn_{\text{A}}n_{\text{B}} - \delta n_0 [\Gamma_{\delta n}+ 2k(n_{\text{A}}+n_{\text{B}})].
\label{eqn:nDifferentialLin} 
\end{equation}
and
\begin{equation} 
\dot\delta n_1 = k(\delta n_0)^2- \delta n_1 [\Gamma_{\delta n}+ 2k(n_{\text{A}}+n_{\text{B}})].
\label{eqn:nDifferential_n1} 
\end{equation}
Both Eq.\,\ref{eqn:nDifferentialLin} and Eq.\,\ref{eqn:nDifferential_n1} can be solved analytically as 
\begin{equation}
    \delta n_0(t)=e^{-q(t)}\int_0^tdt'e^{q(t')}2kn_{\text{A}}(t')n_{\text{B}}(t')
    \label{eqn:nSolved}
\end{equation}
and
\begin{equation} 
\delta n_1(t)=e^{-q(t)}\int_0^tdt'e^{q(t')}k(\delta n_0(t'))^2
\label{eqn:n1Solved} 
\end{equation}
with
\begin{equation}
    q(t)=\int_0^tdt'[\Gamma_{\delta n}+2k\left(n_{\text{A}}(t')+n_{\text{B}}(t'))\right].
    \label{eqn:q}
\end{equation}

\begin{figure}
\centering
\includegraphics[width=1\hsize]{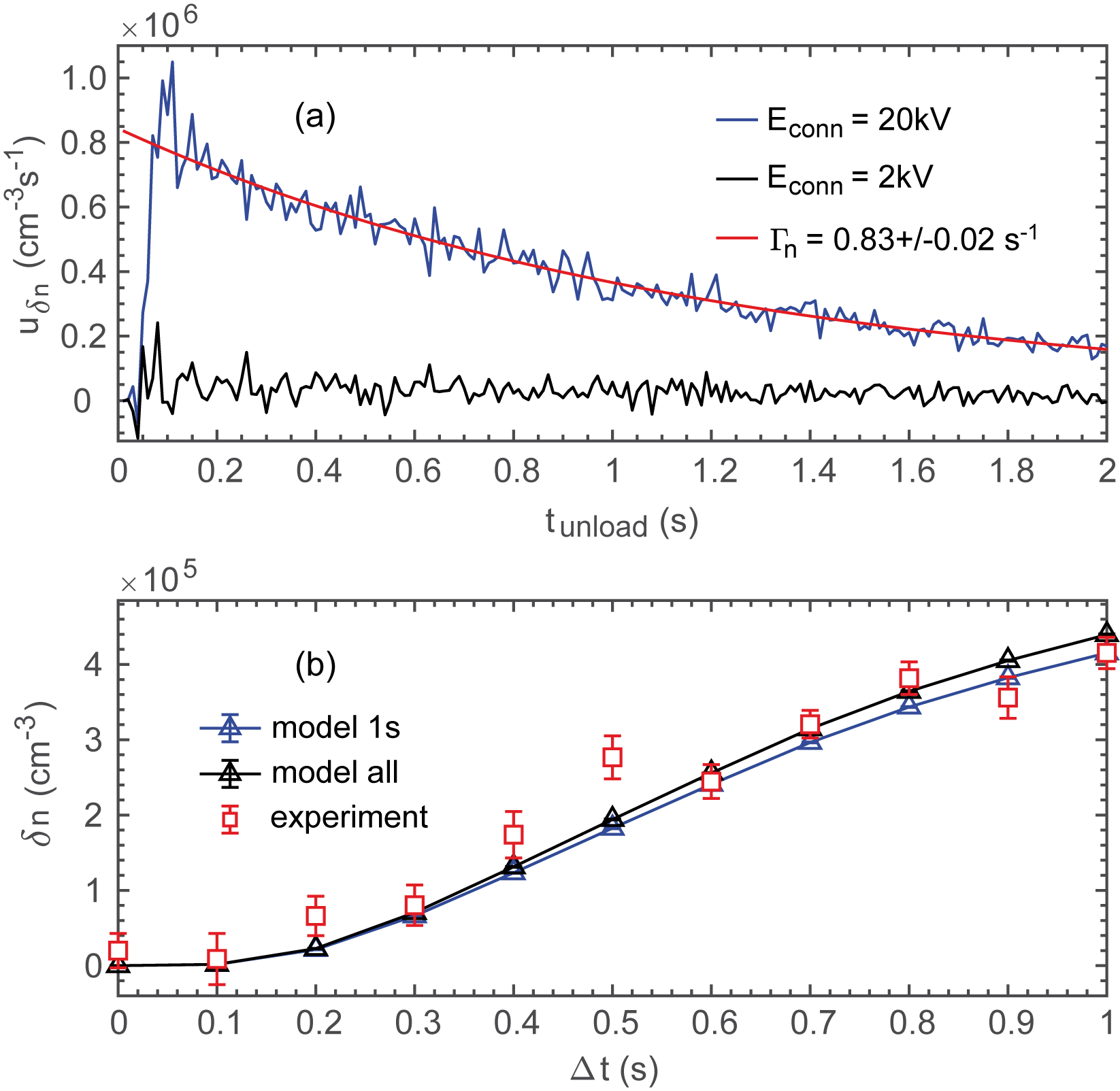}
\caption{Collision signal. (a) Trap unloading signal of the collision data $u_{\delta n}$ as a function of the unloading time $t_{\text{unload}}$ for the connection guide fields $E_{\text{conn}}=20$\,kV/cm (blue) and $E_{\text{conn}}=2$\,kV/cm (black) at a control field of $E_{\text{c}}=0.50$\,kV/cm and an interaction time of $\Delta t=1$\,s. The red line depicts a single-exponential fit to the $E_{\text{conn}}=20$\,kV/cm data. (b) Integrated collision signal ($E_{\text{conn}}=20$\,kV/cm, $E_{\text{c}}=2.37$\,kV/cm) as a function of the interaction period $\Delta t$ (red squares). The black and blue triangles display the solution to the differential Eq.\,\ref{eqn:nDifferentialSuppl} fitted to all measured data and that fitted to the data point at $\Delta t=1$\,s only, respectively. The black and blue solid lines are a guide to the eye.} 
\label{fig:ExtractK}
\end{figure}

This allows us to obtain an expression for the additional loss, $\delta n= \delta n_0 + \delta n_1$, which is a function of the two-body loss-rate coefficient $k$, the single-body loss rate $\Gamma_{\delta n}$ and the density of trapped molecules for the A and the B sample, $n_{\text{A}}$ and $n_{\text{B}}$, respectively. We can measure $n_{\text{A}}$ and $n_{\text{B}}$ as a function of the applied control field $E_{\text{c}}$ and interaction time $\Delta t$, as shown in the main text. Moreover, we can also determine the single-body loss rate associated with the molecules lost due to collisions $\Gamma_{\delta n}$. Therefore we record the trap unloading signals $u_{\text{A}}$, $u_{\text{B}}$ and $u_{\text{A+B}}$ (curves above the dashed areas in Fig.\,\ref{fig:scheme} (a)-(c) of the main text), which are a function of the unloading time $t_{\text{unload}}$ and when integrated proportional to the density of trapped molecules. We then obtain the trap unloading signal of the collisions data as a function of the unloading time according to $u_{\delta n}=u_{\text{A}}+u_{\text{B}}-u_{\text{A+B}}$, which is illustrated in Fig.\,\ref{fig:ExtractK} (a) for a low-density ($E_{\text{conn}}=2$\,kV) and high-density sample ($E_{\text{conn}}=20$\,kV) for an applied control field of $E_{\text{c}}=0.50$\,kV/cm and an interaction time of $\Delta t=1$\,s. As expected, the collision signal of the high-density sample is significantly larger than the collision signal of the low-density sample. We extract the single-body loss rate $\Gamma_{\delta n}$ by fitting a single-exponential function to the trap unloading curve, illustrated by the red line in Fig.\,\ref{fig:ExtractK} (a) for the high-density sample, yielding $\Gamma_{\delta n}=0.83\pm0.02$\,s$^{-1}$. With $k$ now being the only free parameter, we can fit the solution of Eq.\,\ref{eqn:nDifferentialSuppl} to the measured collision signal, as illustrated in Fig.\,\ref{fig:ExtractK} (b) for $E_{\text{c}}=2.37$\,kV/cm and $E_{\text{conn}}=20$\,kV. The red squares display the measured data and the blue triangles the solution of Eq.\,\ref{eqn:nDifferentialSuppl} fitted to the collision signal at $\Delta t=1$\,s. Considering only the linear contribution to Eq.\,\ref{eqn:nDifferentialSuppl} we extract $k=(1.72\pm 0.10)\times 10^{-8}$cm$^3$/s, and when we include the quadratic term we get $k=(1.71\pm 0.10)\times 10^{-8}$cm$^3$/s, showing that its contribution is negligible.

As we need to record data for up to five days to obtain a collision signal with sufficient statistical significance for a single interaction time $\Delta t$, the red squares in Fig.\,\ref{fig:ExtractK} (b) are the only data set, where we recorded the collision signal throughout the entire interaction period $\Delta t \in [0,1]$\,s. For all the other collision measurements presented in this letter we only recorded data for an interaction time of $\Delta t=1$\,s to extract the two-body loss-rate coefficient $k$. This is possible since $k$ is the only free parameter when fitting Eq.\,\ref{eqn:nDifferentialSuppl} to the measured collision data. To quantify if this has a significant impact on the extracted value for $k$, we fit the solution of Eq.\,\ref{eqn:nDifferentialSuppl}, illustrated by the black triangles in Fig.\,\ref{fig:ExtractK} (b), to all measured data points yielding a two-body loss-rate coefficient of $k=(1.83\pm 0.15)\times 10^{-8}$cm$^3$/s. This value overlaps within the errorbars with $k=(1.71\pm 0.10)\times 10^{-8}$cm$^3$/s, the result obtained when only taking the collision signal at $\Delta t=1$\,s (blue triangles in Fig.\,\ref{fig:ExtractK} (b)) into account. This shows that it is justified to extract the two-body loss-rate coefficient $k$ by only considering the measured collision signal at $\Delta t=1$\,s.

\section{Dependence of $k$ on $\Gamma$}

\begin{figure}
\centering
\includegraphics[width=1\hsize]{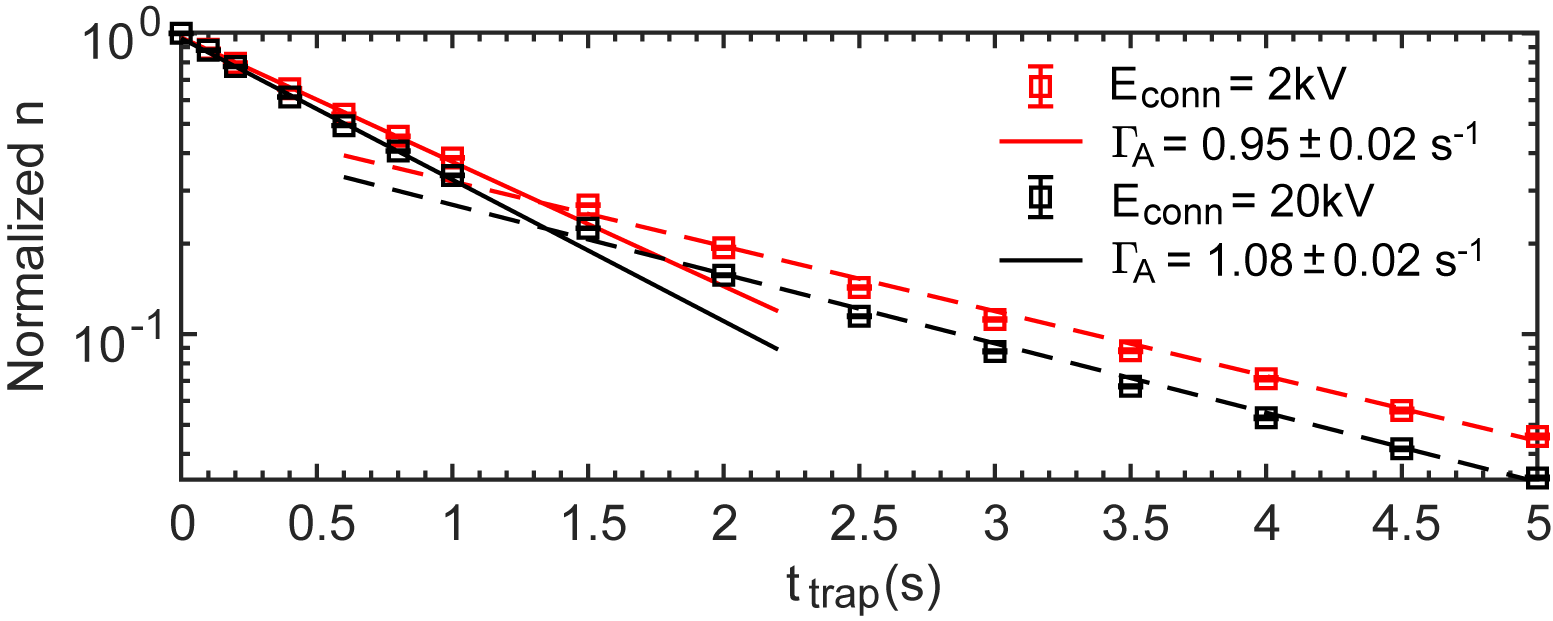}
\caption{Trap lifetime. Normalized trapped density as a function of the trapping time for a low-density ($E_{\text{conn}}=2$\,kV/cm) and high-density ($E_{\text{conn}}=20$\,kV/cm) sample, illustrated by red and black squares, respectively. The solid lines depict a single-exponential fit to the first 1\,s of trapping time, whereas the dashed lines indicate a single-exponential fit to the data for trapping times between 1.5\,s and 5\,s.} 
\label{fig:Lifetime}
\end{figure}

In the main text of this paper we introduced a new measurement scheme to record two-body collisions and extract a precise value for the two-body loss-rate coefficient $k$, insensitive to small changes to the single-body loss rate $\Gamma$. This is important for our system as fast molecules are more prone to get lost from our trap through holes in the potential, like the trap input or exit hole, than slow molecules as they sample these trap regions more frequently. Therefore, $\Gamma$ shows a $v^5$ velocity dependence for molecules with a linear Stark shift causing deviations from a background-limited single-exponential decay in trapped signal for trapping times beyond 1\,s \cite{Zeppenfeld2013}. This is shown in red in Fig.\,\ref{fig:Lifetime} for the A sample with $E_{\text{conn}}=2$\,kV/cm (solid and dashed lines correspond to single-exponential fits to the data within the first 1\,s and within the last 3.5\,s of $t_{\text{trap}}$, respectively), where the corresponding low density is ensuring that collisional trap losses are negligible (see Fig.\,\ref{fig:ExtractK} (a)). Additionally, for a high-density sample ($E_{\text{conn}}=20$\,kV/cm) trap losses due to collisions can be observed, which also cause deviations from a single-exponential decay of the density n as a function of the trapping time $t_{\text{trap}}$ (see black squares in Fig.\,\ref{fig:Lifetime})

In the following, all of the presented single-body loss rates $\Gamma$ are obtained from a fit to the 1\,s of trapping time, to match the interaction period $\Delta t$ for the collision experiments presented in this letter. We observe a clear deviation between the decay rates $\Gamma_{\text{A}}=0.95\pm0.02$\,s$^{-1}$ and $\Gamma_{\text{A}}=1.08\pm0.02$\,s$^{-1}$ for the low- and the high-density sample indicating the observation of collision-induced trap loss. However, to quantitatively determine the impact of collisions we need to determine the two-body loss-rate coefficient $k$.

\begin{figure}
\centering
\includegraphics[width=1\columnwidth]{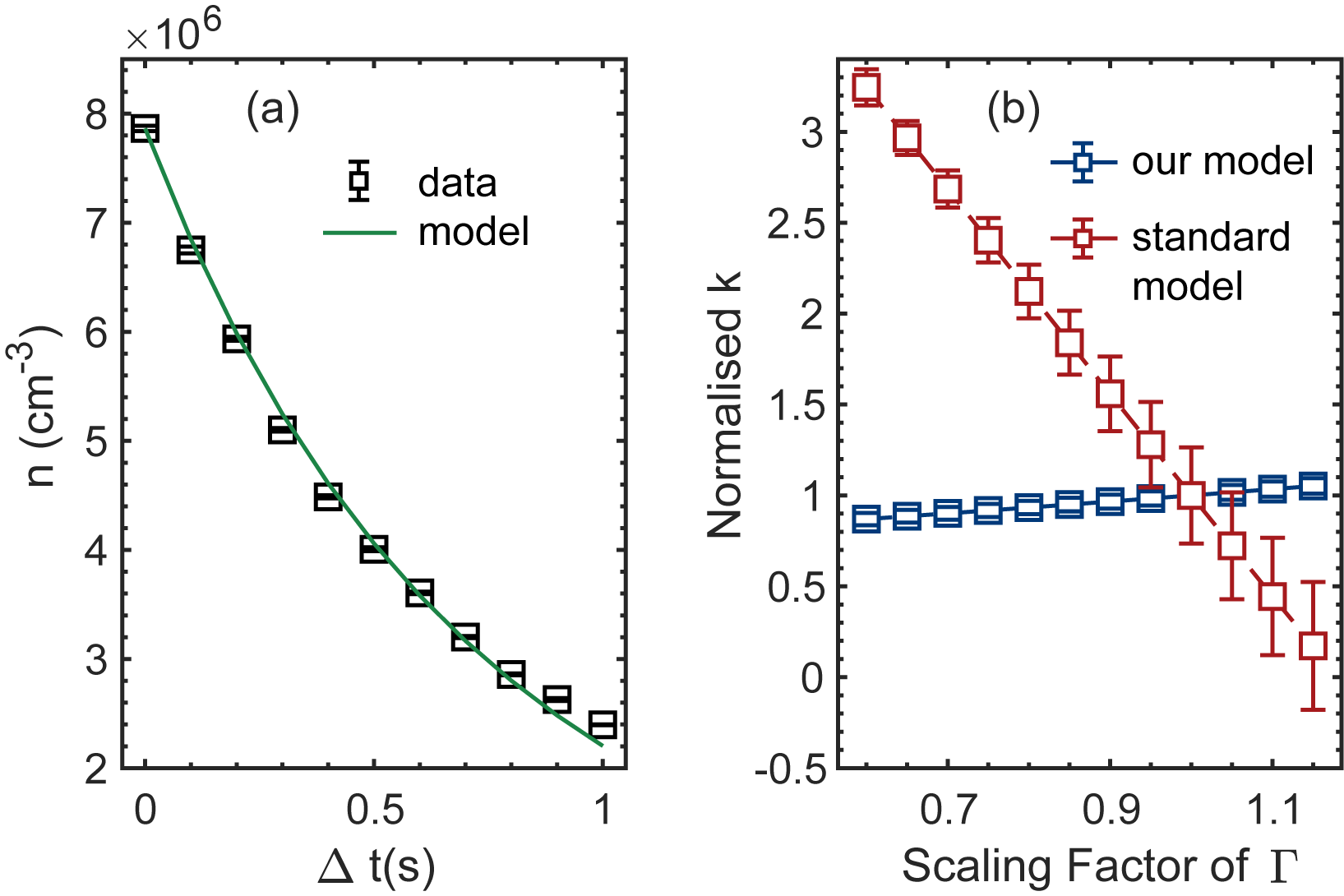}
\caption{Sensitivity of $k$ on the single-body loss rate. (a) Trapped density of the A sample as a function of the interaction time $\Delta t$ (black squares). The green line is a fit of Eq.\,\ref{eqn:nASolved} to the recorded data. (b) Dependence of the two-body loss-rate coefficient $k$ on the single-body loss-rate scaling factor $s$ for the standard method (red) and the measurement scheme developed in this letter (blue).}
\label{fig:kScaling}
\end{figure}

The standard approach \cite{Segev2019} to obtain $k$ is to extract $\Gamma_{\text{A}}$ from the low-density sample and then fit the solution of Eq.\,\ref{eqn:Densities} with $\lambda_{\text{A}}=0$, given by 
\begin{equation} 
    n_{\text{A}}(t)=-\frac{n_{0}\cdot\Gamma_{\text{A}}}{k\cdot n_{0}-(\Gamma_{\text{A}} + k\cdot n_{0})\cdot e^{\Gamma_{\text{A}} \cdot t}},
    \label{eqn:nASolved} 
\end{equation}
to the high-density sample, such that $k$ is the only free parameter. This is illustrated by the green solid line in Fig.\,\ref{fig:kScaling} (a), where the black squares depict the high-density sample, or equivalently the A sample with $\lambda_{\text{A}}=0$ and $E_{\text{conn}}=20$\,kV/cm. However, to use this method in our system, it would have to be insensitive to small changes in the single-body loss rate. This is necessary, as although it is possible to approximate the trap lifetime by a single-exponential decay for holding times shorter than 1\,s (see Figure \ref{fig:Lifetime}), we still observe small deviations from a pure single-exponential decay. In addition, changing the connection guide trapping field from $E_{\text{conn}}=20$\,kV/cm to $E_{\text{conn}}=2$\,kV/cm leads to small changes in the velocity distribution and consequently to changes in the single-body loss rates of the high- and the low-density sample. 

Therefore we compare the sensitivity of $k$ on $\Gamma$ for the measurement scheme introduced in the main text with the standard approach to measure collisions, introduced in the previous paragraph. For the standard approach we multiply $\Gamma_{\text{A}}$ with a scaling factor $s\in[0.60,1.15]$ and fit Eq.\,\ref{eqn:nASolved} to the data, shown in black in Fig.\,\ref{fig:kScaling} (a), to obtain a value for the two-body loss-rate coefficient $k$ as a function of $s$. The result is illustrated by red squares in Fig.\,\ref{fig:kScaling} (b), where we normalized $k$ to the value obtained for the measured single-body loss rate ($s=1$). We perform the identical procedure for the measurement scheme presented in the main text, but here we scale $\Gamma_{\delta n}$ and fit the solution of Eq.\,\ref{eqn:nDifferentialSuppl} to the measured collision signal $\delta n$. The resulting two-body loss rates are displayed by blue squares in Fig.\,\ref{fig:kScaling} (b) showing, in contrast to the data depicted in red, only a small dependence on the scaling factor $s$. We conclude that the measurement scheme developed in this letter is insensitive to small changes in the single-body loss rate and most importantly significantly less sensitive than the standard method and thus well suited to extract a precise value for $k$ in our system.

\section{Calculation of the elastic two-body loss-rate coefficient $k_{\text{el}}$}

\begin{figure}
    \centering
    \includegraphics[width=1\hsize]{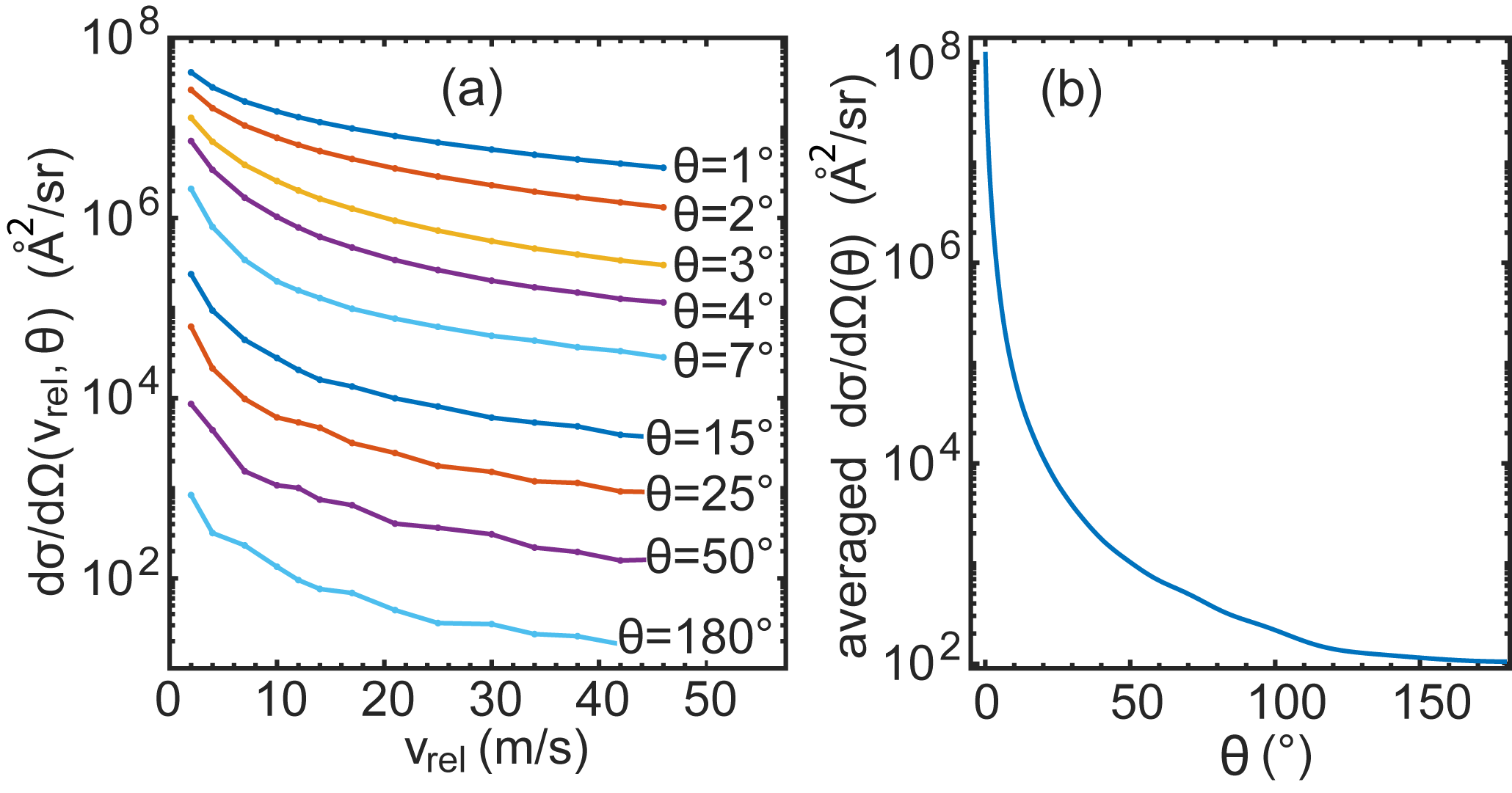}
    \caption{Differential elastic-collision cross-section. (a) Differential cross-section for selected scattering angles $\theta$ as a function of the relative velocity $v_{\text{rel}}$ of the colliding molecules. (b) Differential cross-section averaged over the relative-velocity distribution of the trapped molecules as a function of the scattering angle $\theta$.}
    \label{fig:Eikonal}
\end{figure}

In this section we calculate the two-body loss-rate coefficient $k_{\text{el}}$ due to elastic energy-exchanging collisions. Therefore we have to consider the elastic collision cross-section $\sigma_{\text{loss}}^{\text{el}}(v_{\text{rel}})$ of the trapped molecules and $P_{\text{loss}}(v_{\text{rel}},\theta)$, the probability for a molecule to be lost from the trap in an elastic-collision process \cite{Wu2017}. Both contributions are a function of the relative velocity of the colliding molecules and $P_{\text{loss}}(v_{\text{rel}},\theta)$ in addition depends on the scattering angle $\theta$. As the mean collision energy is large in our system, $\bar{E}_{coll}=k_\text{B}\cdot0.4K$, we compute the differential elastic-collision cross-section $\frac{d\sigma}{d\Omega}(v_{\text{rel}},\theta)$ using the semi-classical eikonal approximation \cite{Bohn2009, Wu2017}. Taking into account the isotropic part of the dipole-dipole interaction, we obtain $\frac{d\sigma}{d\Omega}(v_{\text{rel}},\theta)$, as illustrated in Fig.\,\ref{fig:Eikonal} (a) for selected scattering angles $\theta\in[0,180^\circ]$, as a function of $v_{\text{rel}}$. The semi-classical nature of the collision process, strongly favoring forward scattering, is clearly visible in Fig.\,\ref{fig:Eikonal} (b), displaying the differential elastic cross-section averaged over the relative-velocity distribution in our trap.

We integrate the differential cross-section $\frac{d\sigma}{d\Omega}(v_{\text{rel}},\theta)$ over the full solid angle $4\pi$ to obtain the elastic collision cross-section as a function of $v_{\text{rel}}$   
\begin{equation}
    \sigma_{\text{el}}(v_{\text{rel}})=\int\frac{d\sigma}{d\Omega}(v_{\text{rel}},\theta)d\Omega,
    \label{eqn:SigmaElastic}
\end{equation}
which we average over the relative-velocity distribution of the colliding molecules to arrive at $\sigma_{\text{el}}=6.97\times10^{-12}$cm$^2$ and $\sigma_{\text{el}}=7.50\times10^{-12}$cm$^2$ for $E_{\text{c}}=0.50$\,kV/cm and $E_{\text{c}}=2.37$\,kV/cm, respectively. We want to note that the elastic collision process itself does not show an electric-field dependence, however a change in the control field alters the relative-velocity distribution and consequently the elastic cross-section $\sigma_{\text{el}}$.
\begin{figure}
    \centering
    \includegraphics[width=1\hsize]{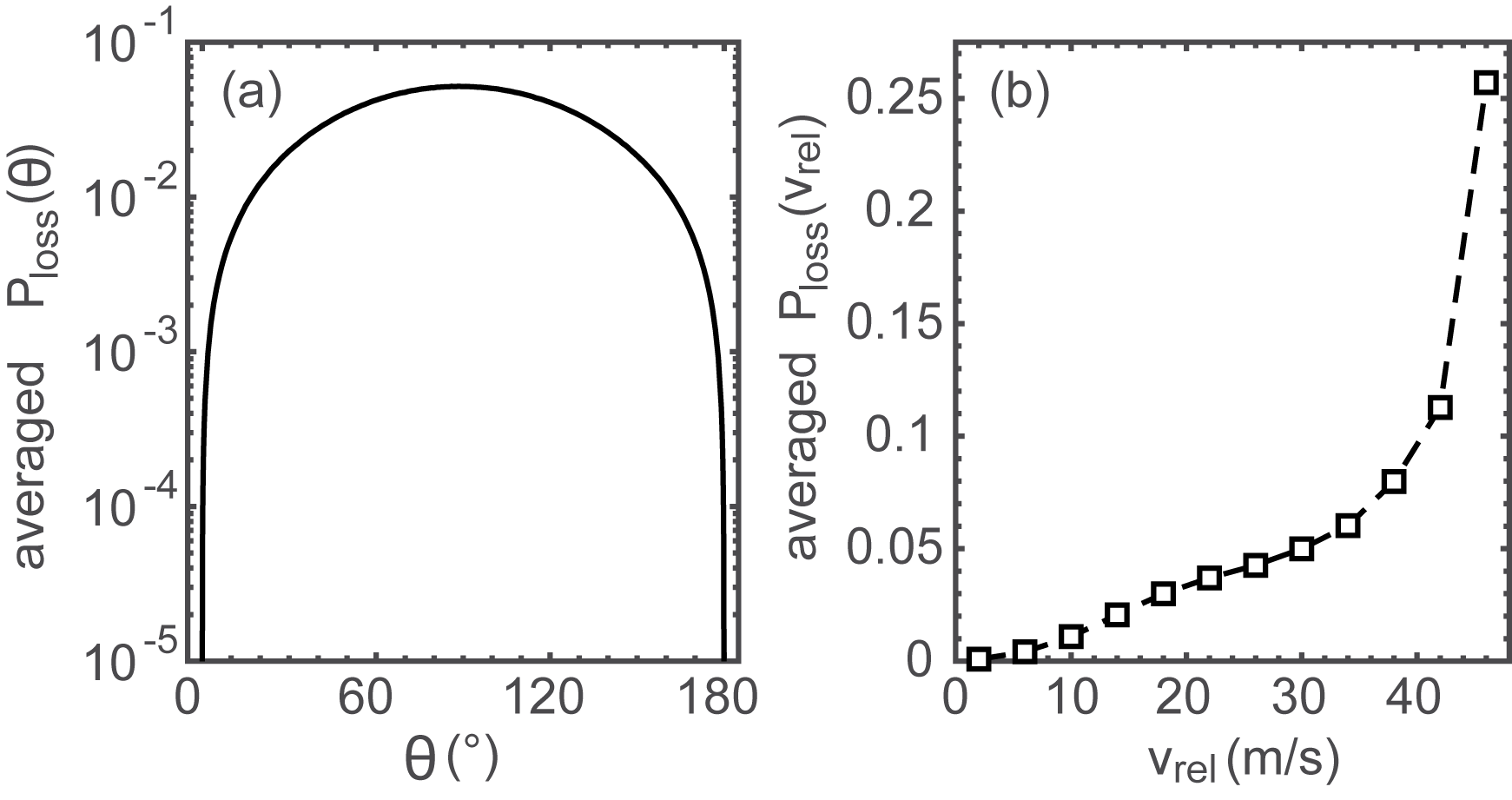}
    \caption{Loss probability. (a) Likelihood for a a molecule to be lost from the trap due to an elastic-collision process, averaged over the molecules' relative-velocity distribution in the trap, as a function of the scattering angle $\theta$. (b) $P_{\text{loss}}(v_{\text{rel}})$, averaged over the scattering angle $\theta$, plotted against the relative velocity $v_{\text{rel}}$.}
    \label{fig:Ploss}
\end{figure}

In order to determine the elastic loss cross-section $\sigma_{\text{loss}}^{\text{el}}(v_{\text{rel}})$ we also need to consider $P_{\text{loss}}(v_{\text{rel}},\theta)$, the likelihood for a molecule to be lost in an elastic collision event. Therefore we utilize Monte-Carlo simulations, taking into account the electric-field distribution in the trap and the molecules' relative-velocity distribution. For a given scattering angle $\theta$, sampled from a flat distribution between $0^\circ$ and $180^\circ$, we compute the energy transfer in an elastic-collision process in the center-of-mass-frame. If the molecule's total energy after the collision process in the laboratory frame exceeds the trap depth we count the molecule as lost. In addition we consider a second contribution to $P_{\text{loss}}(v_{\text{rel}},\theta)$ resulting from collision-induced changes to the energy-dependent single-body loss rate in our electric trap. To account for this we compare the molecules' kinetic energy before and after the collision process in the laboratory frame and determine the resulting change in the likelihood for the molecules to be lost from the trap. Taking both contributions into account Fig.\,\ref{fig:Ploss} (a) illustrates $P_{\text{loss}}(\theta)$, averaged over the relative-velocity distribution in the trap, as a function of the scattering angle $\theta$. We observe a maximum in the likelihood for a loss to occur at a scattering angle of $90^\circ$ and a steep decrease in $P_{\text{loss}}(\theta)$ for small and large scattering angles. Besides the angular dependence, we can also inspect the loss probability as a function of the relative velocity of the colliding molecules, depicted in Fig.\,\ref{fig:Ploss} (b), where a larger relative velocity is more likely to lead to an elastic-collision-induced trap loss.

With this in place we can determine the elastic-loss cross-section as a function of the relative velocity according to 
\begin{equation}
    \sigma_{\text{loss}}^{\text{el}}(v_{\text{rel}})=\int\frac{d\sigma}{d\Omega}(v_{\text{rel}},\theta)P_{\text{loss}}(v_{\text{rel}},\theta)d\Omega,
    \label{eqn:SigmaLossElastic}
\end{equation}
and are able to calculate the corresponding elastic two-body loss-rate coefficient as $k_{\text{el}}(v_{\text{rel}})=\sigma_{\text{loss}}^{\text{el}}(v_{\text{rel}}) v_{\text{rel}}$. Averaging over the relative-velocity distribution in the trap allows us to determine $k_{\text{el}}$ for a given control field, e.g. $k_{\text{el}}=4.51\times10^{-11}$cm$^3$/s for $E_{\text{c}}=2.37$\,kV/cm. Again, as already mentioned in this section, the elastic-collision process does not show an explicit electric-field dependence. However, changes in the relative-velocity distribution for different control fields slightly affect $\sigma_{\text{loss}}^{\text{el}}$. In total, we can summarise that the elastic cross-section itself is big, however it is unlikely that a molecule is lost from the trap due to an elastic collision process and the associated two-body loss-rate coefficient $k_{\text{el}}$ is roughly three orders of magnitude smaller than the two-body loss-rate coefficient due to dipolar relaxation, as shown in the main text of this letter.

\section{Error budget of the measured two-body loss-rate coefficient $k$}
The error bars for the measured two-body loss-rate coefficient $k$ contain statistical and systematic uncertainties. The main error source in our experiment are statistical errors due to fluctuations of the molecule signal, which are a consequence of temperature changes in the buffer-gas cell. These temperature changes are a result of the duty cycle of our pulse tube cooler (Cryomech PT420 pulse). The fluctuations in molecule signal directly affect the trapped densities of the A, B and A+B samples and thereby also the collision signal $\delta n$ and the two-body loss-rate coefficient $k$. The second contribution to the statistical error budget arises from the single-exponential fit to extract the single-body loss rate $\Gamma_{\delta n}$, exemplarily shown in Fig. \ref{fig:ExtractK} (a). The respective errors of the measured densities $\epsilon n_{\text{A}}(t)$, $\epsilon n_{\text{B}}(t)$, of the lifetime $\epsilon \Gamma_{\delta n}$, and of the measured collision signal $\epsilon \delta n(t)$, are propagated to yield an error $\epsilon A(k,t)$ on the parametric expression 
\begin{equation}
    A(k,t):= \delta n(t) - \delta n_0(k,t)- \delta n_1(k,t)=0, 
    \label{eqn:Error_k}
\end{equation}
where $\delta n(t)$ is the measured collision signal and $\delta n_0(k,t)$ and $\delta n_1(k,t)$ are given by Eq.\,\ref{eqn:nSolved} and Eq.\,\ref{eqn:n1Solved}, respectively. By finding the most appropriate values $k_{\text{up}}$, $k$ and $k_{\text{low}}$ that solve the three equations
\begin{align}
\begin{split}
    A(k_{\text{up}},t)+\epsilon A(k_{\text{up}},t)=0\\
    A(k,t)=0\\
    A(k_{\text{low}},t)-\epsilon A(k_{\text{low}},t)=0,\\
\end{split}
\label{eqn:error_k}
\end{align}
we obtain an upper bound ($k_{\text{up}}$), an estimate ($k$) and a lower bound ($k_{\text{low}}$) for the two-body loss-rate coefficient, respectively. The larger difference between $|k_{\text{up}}-k|$ and $|k-k_{\text{low}}|$ is defined as the confidence interval of $k$ and is plotted as symmetric error bars on $k$. Systematic effects are caused by uncertainties in our QMS density calibration, which affect the nominal density of the trapped ensembles and thereby the collision signal $\delta n$ and the two-body loss-rate coefficient $k$, creating a global scaling of up to a factor of four.

\end{document}